\documentclass[9pt, twocolumn]{article}
\usepackage{textgreek}
\usepackage[utf8]{inputenc}
\usepackage{graphicx}
\graphicspath{Figures}
\usepackage[
    a4paper,
    left=15mm,
    right=15mm,
    top=18mm,
    bottom=18mm
]{geometry}
\usepackage{fancyhdr}
\usepackage{amsmath}
\usepackage{caption}
\usepackage{subcaption}
\usepackage{url}
\usepackage[hidelinks]{hyperref}
\usepackage{cleveref}
\usepackage{lscape}
\usepackage{mathtools}
\usepackage{xcolor}
\usepackage{pdflscape}
\usepackage{multirow}
\usepackage{float}
\usepackage{pifont}
\usepackage{longtable}
\usepackage{multicol}
\usepackage{booktabs}
\usepackage{siunitx}
\usepackage{adjustbox}
\usepackage{abstract}
\everymath{\displaystyle}

\crefname{equation}{Equation}{Equations}
\Crefname{equation}{Equation}{Equations}

\begin{document}

\twocolumn[
\begin{@twocolumnfalse}

\title{ Geometry-Controlled Dynamic Tensiometry Resolves Intrinsic Surfactant Adsorption Kinetics}
\author{Camille Brigodiot\(^{a}\)\footnote{Corresponding author. \\E-mail adress: c.brigodiot@uva.nl}, Boxin Deng\(^b\), Christine Dalmazzone\(^c\), Karin Schroën\(^b\), Annie Colin\(^d\)\\
\and a: Van der Waals - Zeeman Institute, Institute of Physics, University of Amsterdam,\\ Science Park 904, 1098 XH Amsterdam, The Netherlands \\
b: Wageningen University and Research, Laboratory of Food Process Engineering,\\ Bornse Weilanden 9, 6708 WG Wageningen, the Netherlands \\
c: 	IFP Energies nouvelles, 1 et 4 avenue de Bois-Préau, 92852 Rueil-Malmaison, France \\
d: MIE, CBI, ESPCI Paris, Université PSL, CNRS 75005 Paris, France }
\date{}
\maketitle


\begin{abstract}
At short times, interfacial tension depends on experimental geometry  because surfactant transport to the interface depends on the  mass-transfer conditions.A predictive description therefore requires  more than a dynamic tension curve or a fitted adsorption constant:  interfacial thermodynamics, diffusion, and adsorption kinetics must be  identified separately.
Here, we combine equilibrium and diffusion measurements with a microfluidic EDGE tensiometer that provides a nearly stationary interface and controlled micrometer-scale transport. Equilibrium properties and diffusion are determined independently, leaving adsorption kinetics as the key unknown. Dynamic tension is then calculated using a nonequilibrium thermodynamic description, without assuming instantaneous equilibrium between the adsorbed layer and the subsurface solution.
For the nonionic surfactant C$_{10}$E$_{8}$, equilibrium thermodynamics and transport are independently constrained, and a single intrinsic adsorption rate constant describes several concentrations. We extend the framework to SDS by including electrostatic interactions and subsurface-concentration dynamics,  capturing transient depletion and replenishment. Once thermodynamic, transport, and kinetic parameters are identified, the model predicts dynamic interfacial tension beyond the geometry and conditions used to determine them.
The microfluidic EDGE tensiometer thus provides both a reliable short-time tensiometry method and a quantitative framework for identifying the physical mechanisms governing surfactant mass transfer at interfaces.
\end{abstract}

\end{@twocolumnfalse}
]

\section{Introduction}

Dynamic interfacial tension controls droplet and bubble formation in emulsification, foaming, spraying, coating, and many pharmaceutical, cosmetic, and food-processing operations \cite{maan2011spontaneous,schultz2004high,malysa2005influence}. In these processes, interfaces are often created within milliseconds, long before surfactants reach equilibrium. Their behavior therefore depends on how rapidly surfactants are transported through the liquid and transferred to the newly created interface.

Accessing short interface ages is not sufficient to determine these kinetics. The measured tension relaxation combines bulk transport and molecular adsorption. An intrinsic adsorption rate constant can be identified only when bulk transport is independently constrained or is effectively faster than interfacial transfer. Otherwise, different combinations of diffusion and adsorption parameters may produce similar tension--time curves. The challenge is therefore to probe newly formed interfaces in a geometry with a well-defined rate-limiting mechanism \cite{kovalchuk2023surfactant}.

Classical pendant- and rising-drop tensiometers provide accurate liquid--liquid measurements but involve transport over macroscopic distances. At dilute concentrations, their long-time response is commonly diffusion-controlled \cite{lin1990diffusion,berry2015measurement,nele2022analytically}.
These measurements can determine the diffusion coefficient, but become insensitive to molecular adsorption when interfacial transfer is faster than bulk replenishment. Maximum-bubble-pressure tensiometry reaches millisecond surface ages \cite{garrett1989reexamination,christov2006maximum,mishchuk2001hydrodynamic}, but the interface continuously expands and the measured pressure also depends on bubble hydrodynamics, surface dilution, and the definition of surface age. Shorter times therefore do not necessarily provide direct access to adsorption kinetics.

Microfluidics offers a promising alternative because small dimensions and convective renewal can accelerate bulk transport. Alvarez et al. developed a microtensiometer in which flow is imposed around a spherical air--water interface held at the end of a glass capillary \cite{Alvarez.2010micro, alvarez2012using, reichert2015importance}. Reducing the interface radius and the convective boundary-layer thickness was intended to make adsorption rate-limiting. For C$_{12}$E$_8$ and C$_{14}$E$_8$, however, the measured dynamics remained controlled by convection--diffusion, and only lower bounds on the adsorption constants could be obtained.

Brosseau, Vrignon, and Baret used the same timescale-separation principle with convected microdroplets \cite{Brosseau.2014, Riechers.2016}. Water droplets were transported through a delay line containing repeated narrow channels and planar expansions. Their age was set by the travel time, and the interfacial tension was inferred from their deformation at each expansion. For the PEG--PFPE surfactant investigated, transport was estimated to be faster than molecular transfer, allowing adsorption and desorption constants to be extracted. The tension, however, was reconstructed from the deformation of a confined moving droplet through a calibrated hydrodynamic relation, rather than directly measured.

Differential-pressure measurements during microfluidic droplet formation provide another route to short-time interfacial tension \cite{Liang.2022}. In this geometry, the droplet radius, interfacial area, surface-dilution rate, and surrounding flow evolve simultaneously. The dominant mechanism must therefore be identified a posteriori from operating conditions and model comparison, and the inclusion of an adsorption constant does not ensure that it is experimentally identifiable.

A geometry is thus still needed that combines direct oil/water tension measurement, short interface ages, controlled subsurface renewal, and direct assessment of the mechanism governing the response. Here, we introduce a microfluidic EDGE (Edge-based Droplet GEneration) tensiometer designed to meet these requirements \cite{Deng2022EDGE}. An oil meniscus is confined inside a shallow pore and exposed to a continuously renewed aqueous surfactant solution. During most of the adsorption period, the meniscus remains pinned and its geometry varies only weakly. Adsorption lowers the interfacial tension and the Laplace pressure retaining the meniscus. When the applied pressure exceeds this capillary threshold, the interface advances and a droplet is released. The waiting time defines the interface age, while the Young--Laplace relation gives the corresponding tension.

We first compare the characteristic transport and adsorption times of the principal tensiometric geometries. This analysis reveals the advantages of EDGE: a short and well-defined transport distance, continuous subsurface renewal, and a nearly stationary interface with weak variations in area and curvature. It establishes the conditions under which the EDGE response is sensitive to molecular adsorption rather than dominated by bulk transport.

The kinetic parameters are then determined sequentially. Equilibrium tensiometry provides the adsorption isotherm and interfacial equation of state, while dilute pendant-drop measurements in a diffusion-controlled regime determine the diffusion coefficient \(D\). This independently measured value is fixed in the EDGE analysis. Adsorption-limited, diffusion-limited, and mixed models are then confronted with the EDGE interfacial-tension measurements to identify the adsorption and desorption parameters without compensating for an unknown bulk-transport rate.

We apply this strategy to the nonionic surfactant C$_{10}$E$_8$ and sodium dodecyl sulfate (SDS) at the water--hexadecane interface. The kinetic parameters are then determined sequentially. 

\subsection{Microfluidic EDGE tensiometer}

\subsubsection{Microchip and set-up}

\begin{figure*}[ht]
    \centering
    \includegraphics[width=15cm]{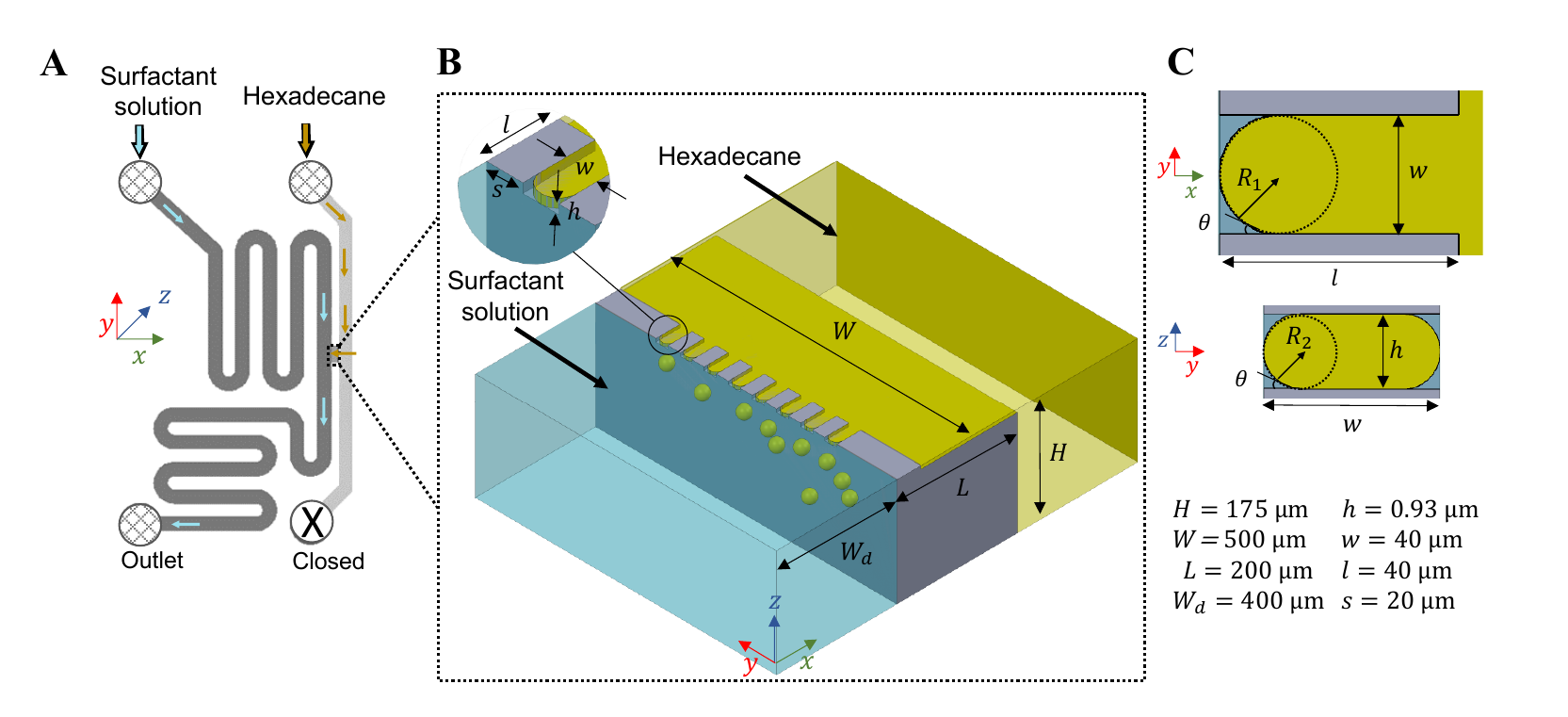}
    \caption{Schematic illustration of the partitioned-EDGE device (A) and of the plateau and the pores (B). Side schematic views of curvatures in the two different directions in a pore (C). Adapted from Santos et al. \cite{santos2024interfacial}.}
    \label{fig:EDGE2}
\end{figure*}

The partitioned-Edge-based Droplet GEneration (EDGE) microchips have been produced by Micronit Microtechnologies B.V (Enschede, The Netherlands), and they have been previously used by Deng et al. \cite{Deng.2022} and Santos et al. \cite{santos2024interfacial} as microfluidic tensiometers. The partitioned-EDGE microchip consists of two deep channels, one straight channel for the dispersed phase and one meandering channel for the continuous phase (see \autoref{fig:EDGE2}A). The deep channels with 175 $\mu m$ height ($H$) and 400 $\mu m$ ($W_d$) of width are connected in the parallel-running section by a shallow plateau of 200 $\mu m$ in length ($L$) and 500 \(\mu m\) in width ($W$). The shallow plateau is further partitioned into eight identical parallel pores; these pores are 40 $\mu m$ in length ($l$) and have a rectangular cross-section with 40 $\mu m$ width ($w$). The plateau and pores have a height of h =  0.93 $\mu m$. The partition width ($s$) is 20 $\mu m$.

The chip was set in a chip holder (Micronit B.V., The Netherlands) and connected via PEEK tubing (0.75 mm, BGB, Switzerland) to a digital pressure controller operated through Smart Interface software (Elveflow, France). The pressure of the continuous phase \(P_c\) was kept constant at 100 $mbar$, while that of the dispersed phase \(P_d\) was modified manually, after closing its outlet (\autoref{fig:EDGE2}A).

By keeping \(P_c\) constant, any effects related to variations in shear flow are avoided. Based on the distribution of flow resistance in the system, the effective pressure \( P_d^{\ast}\) across the plateau and the pores is estimated as: $ P_d^{\ast} = P_d - P_c/2$. Interfacial tension $\gamma$ is calculated by the following equation \cite{cao2026heat}:

\begin{equation}
    \gamma = \frac{P^\ast_d}{\frac{1}{R_1} + \frac{1}{R_2} } = \frac{P^\ast_d}{2} \Big(\frac{wh}{(w+h)} \mathrm{cos(\theta)} \Big)
\end{equation}

where $R_1$ and $R_2$ are the principal radii of curvature for the top and front view, respectively (\autoref{fig:EDGE2}). A value of $\sim 30$° was measured for $\theta$ and used here.

\subsubsection{Surfactant transport and adsorption at the EDGE meniscus}

We now examine how the EDGE geometry controls surfactant supply to the interface by combining continuous convective renewal above the pore with diffusion towards the recessed and nearly stationary meniscus. Surfactant transfer occurs through three successive steps. First, the flowing aqueous phase continuously renews the solution above the pore and limits the growth of the local depletion layer. Second, surfactant molecules diffuse through this depletion layer and across the recessed region separating the renewed liquid from the oil--water meniscus. Third, once they reach the meniscus, the surfactant molecules adsorb at the interface.

The first two steps determine the rate at which the surfactant is supplied to the interface. Their combined transport resistance is represented by an
effective transport length defined as the following (see Supporting Information Fig.S1 and S2 for details):

\begin{equation}
L_D = \frac{w}{ \displaystyle
\int_{-w/2}^{w/2} \frac{\mathrm{d}y}{ \left[\pi D(y+w/2)/U\right]^{1/2}
+ \ell_m(2y/w)^2}}
\label{eq:EDGE_transport_length}
\end{equation}

where $U$ is the effective renewal velocity, $w$ is the pore width, $D$ is the surfactant diffusion coefficient, and $\ell_m$ is the maximum recession length of the meniscus (Fig. S1). For fixed flow conditions and meniscus geometry, $L_D$ remains approximately constant during the measurement and can be calculated independently of the dynamic interfacial-tension curves. Increasing the renewal velocity reduces the contribution of the depletion layer to $L_D$ and therefore increases the
surfactant supply rate.

The third step is the molecular adsorption of the surfactant at the oil--water interface. The progressive accumulation of adsorbed molecules lowers the interfacial tension and consequently decreases the Laplace pressure retaining the meniscus inside the pore. When the Laplace pressure becomes equal and exceeds the imposed effective pressure due to surfactant adsorption, the meniscus advances and a droplet is released. The droplet-formation time therefore measures the time required for transport and adsorption to reduce the interfacial tension to the selected threshold value. The regime analysis introduced below is used to determine whether this measured time is primarily controlled by bulk supply or by molecular adsorption.

Because droplet formation is highly periodic, the initial state of the meniscus is reproducible from one cycle to the next for a given concentration and applied pressure. 
Further experimental details are provided in \autoref{sec:Mat_Met}, including the operating conditions of the EDGE device, the data treatment and the experimental limits of the technique.

\section{Geometry and characteristic kinetic times}

\begin{figure*}[ht]
    \centering
    \includegraphics[width=18cm]{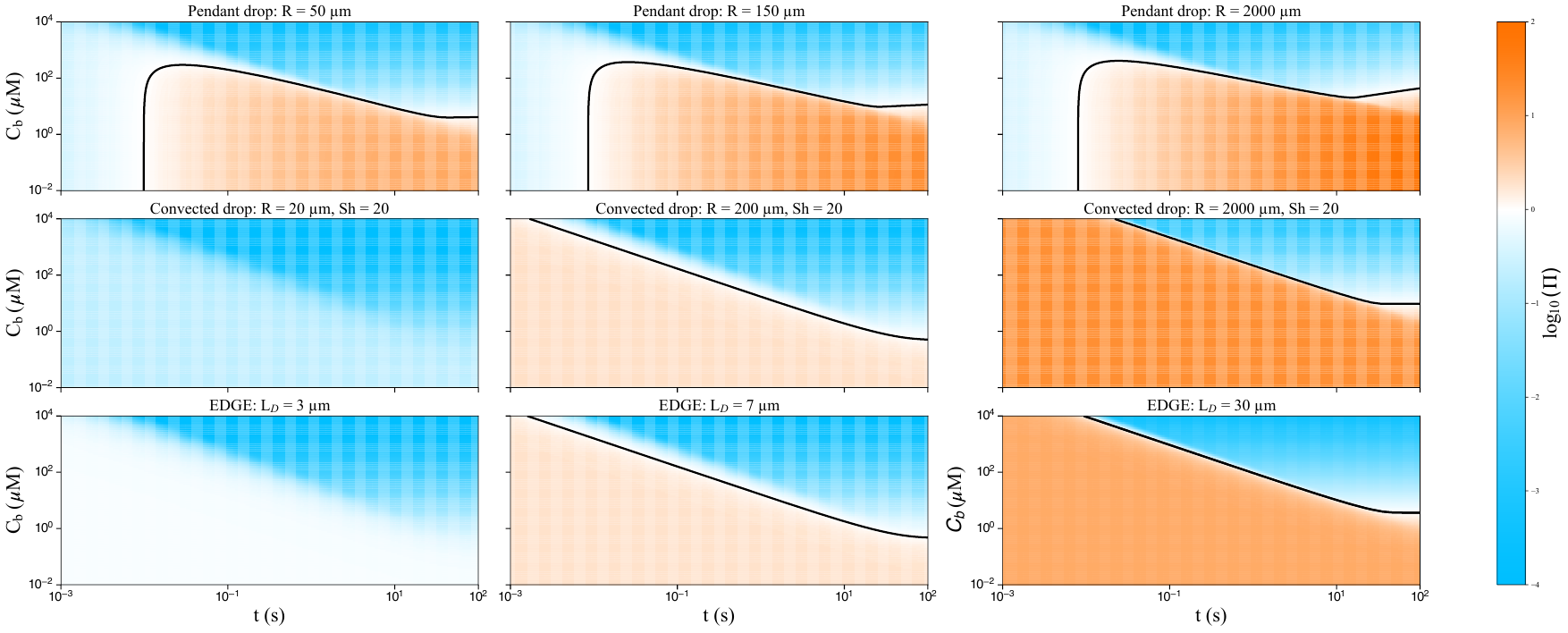}
    \caption{\textbf{Regime maps for pendant drops, convected drops, and the EDGE geometry.} The maps show $\log_{10}(\Pi)$ as a function of interface age $t$ and bulk surfactant concentration $C_b$, with $\Pi=\Gamma_{\infty}k_a(1-\phi)/k_m$. The black contour corresponds to $\Pi=1$ and separates the adsorption-sensitive regime, $\Pi<1$ (blue--cyan), from the transport-sensitive regime, $\Pi>1$ (orange).
    Surface coverage is calculated from \autoref{eq:dphi_dt_adsorption}, coupled to the interfacial mass balance given by \autoref{eq:km_mass_balance}. The upper row corresponds to pendant drops with $R=50$, $150$, and $2000~\mu\mathrm{m}$, from left to right, using $k_m=\sqrt{D/(\pi t)}+D/R$. The middle row corresponds to convected drops with $R=20$, $200$, and $2000~\mu\mathrm{m}$ and $\mathrm{Sh}=20$, using $k_m=\mathrm{Sh}D/R$.
    The lower row corresponds to EDGE geometries with $L_D=3$, $7$, and $30~\mu\mathrm{m}$, using $k_m=(A_f/A_i)D/L_D$, with $A_f/A_i=1/1.38$ (Calculations can be found in Supporting Information S1 and S2). All maps were calculated over $10^{-3}\leq t\leq10^{2}~\mathrm{s}$ and $10^{-2}\leq C_b\leq10^{4}~\mu\mathrm{M}$, with $D=1.0\times10^{-9}~\mathrm{m^2\,s^{-1}}$, $\Gamma_{\infty}=2.0\times10^{-6}~\mathrm{mol\,m^{-2}}$, $K=2000~\mathrm{m^3\,mol^{-1}}$, $k_a=100~\mathrm{m^3\,mol^{-1}\,s^{-1}}$, and $k_d=k_a/K=0.05~\mathrm{s^{-1}}$.
    For pendant drops, the decrease of $k_m$ with interface age can produce re-entrant trajectories between adsorption-sensitive and transport-sensitive regimes.
    By contrast, $k_m$ remains constant at fixed operating conditions for convected drops and EDGE, so that a system initially in the adsorption-sensitive regime tends to remain there as surface coverage increases.}
    \label{fig:regime_maps}
\end{figure*}

To determine whether the measured interfacial dynamics are controlled by molecular adsorption or by bulk transport, we coupled Langmuir adsorption kinetics with a geometry-dependent mass-transfer model. We compared three different geometries: pendant drop, convected drop, and EDGE by constructing the regime maps shown in \autoref{fig:regime_maps}. 
The surface coverage $(\phi)$ evolves according to a Langmuir kinetics as the following expression:

\begin{equation}
\frac{d\phi}{dt}=k_a c_s(1-\phi)-k_d\phi
\label{eq:dphi_dt_adsorption}
\end{equation}

where $c_s$ is the surfactant concentration in the subsurface, and is not assumed to remain equal to the bulk concentration. It is obtained from the following interfacial mass balance: 

\begin{equation} 
k_m(c_b-c_s)=\Gamma_\infty\frac{d\phi}{dt}
\label{eq:km_mass_balance}
\end{equation}

The mass-transfer coefficient $k_m$ accounts for the geometry and hydrodynamic conditions of each experiment. We used $k_m=\sqrt{D/(\pi t)}+D/R)$ for a pendant drop, $(k_m=\mathrm{Sh}D/(2R))$ for a convected drop. The Sherwood number $Sh$ was estimated from the Ranz–Marshall correlation, $\mathrm{Sh}=2+0.6 \,\mathrm{Re}^{1/2}\mathrm{Sc}^{1/3}$.
Here, $(\mathrm{Re}=\rho U_{\mathrm{rel}}(2R)/\mu)$ is the Reynolds number based on the relative velocity $(U_{\mathrm{rel}})$ between the drop and the surrounding liquid, and $(\mathrm{Sc}=\mu/(\rho D))$ is the Schmidt number. 
For the EDGE geometry, we used $k_m=(A_f/A_i)D/L_D$, where $A_f$ is the area of the renewed liquid film, $A_i$ is the interfacial area, and $L_D$ is the effective transport length. 

To compare the relative contributions of interfacial adsorption and bulk mass transfer, we defined the dimensionless ratio: $\Pi=\Gamma_{\infty}k_a(1-\phi)/k_m$.
Here, $\Gamma_\infty$ is the maximum surface excess, $k_a$ is the adsorption rate constant, $\phi$ is the fractional surface coverage, and $k_m$ is the mass-transfer coefficient. The ratio $\Pi$ compares the characteristic rates of interfacial adsorption and bulk mass transfer: $\Pi<1$ indicates that bulk transport is sufficiently fast for the measured dynamics to remain sensitive to the intrinsic adsorption rate, whereas $\Pi>1$ indicates that surfactant supply limits the response.
For each geometry, $\Pi$ was calculated as a function of bulk concentration and interface age to construct the corresponding regime maps. The coupled equations were integrated numerically for each bulk concentration, interface age, and geometry.

\subsection{Interpretation of the regime maps}

\autoref{fig:regime_maps} compares the adsorption- and transport-sensitive regimes obtained for pendant drops, convected drops, and the EDGE geometry. The black contour corresponds to $\Pi=1$ and separates the adsorption-sensitive domain ($\Pi<1$), shown in blue--cyan, from the transport-sensitive domain ($\Pi>1$), shown in orange. At fixed geometry, the regime evolves with bulk concentration and interface age as the surface coverage increases, while the influence of the experimental geometry and hydrodynamic conditions is captured by $k_m$. These maps therefore identify the experimental geometries and operating conditions under which the measured dynamics remain sensitive to the intrinsic adsorption rate constant rather than being limited by bulk transport.

For pendant drops, shown in the upper row, the mass-transfer coefficient is $k_m = \sqrt{{D}/{\pi t}}+{D}/{R}$. The first contribution decreases continuously with interface age because the diffusion layer grows around the drop. Consequently, the transport resistance increases with time, and a system that is initially adsorption-sensitive may progressively become transport-sensitive. Increasing the drop radius from $50$ to $2000~\mu\mathrm{m}$ also decreases the contribution $D/R$ and therefore expands the transport-controlled domain. Some trajectories (for $\mathrm{c_b}$ high enough) cross the regime boundary twice: the system initially transitions from adsorption-sensitive to transport-limited dynamics as $k_m$ decreases, before returning to the adsorption-sensitive regime at long times as progressive interfacial saturation strongly reduces the $(1-\phi)$ factor in $\Pi$. Classical pendant-drop tensiometry is thus particularly sensitive to bulk diffusion for large drops, at long times, and at dilute concentrations.

For convected drops, shown in the middle row, the mass-transfer coefficient is approximated by $k_m =(\mathrm{Sh}D)/(2R).$
At fixed flow conditions, $k_m$ is independent of interface age. Decreasing the drop radius or increasing the Sherwood number enhances mass transfer and shifts the system toward the adsorption-sensitive domain. Conversely, increasing $R$ decreases $k_m$ and favors transport control. Because $k_m$ remains constant while $\phi$ increases during adsorption, the quantity $\Gamma_{\infty}k_a(1-\phi)$ decreases with time. A convected-drop experiment that starts in the adsorption-sensitive regime therefore remains in this regime throughout the relaxation.

For EDGE, shown in the lower row, the mass-transfer coefficient is
$k_m=(A_f/A_i)D/L_D$ where $A_f$ is the area of the renewed liquid film, $A_i$ is the interfacial area, and $L_D$ is the effective transport length. Increasing $L_D$ from $3$ to $30~\mu\mathrm{m}$ decreases the surfactant supply rate and progressively enlarges the transport-sensitive domain. Conversely, reducing $L_D$ shifts complete relaxation toward adsorption control. As in the convected-drop geometry, $k_m$ remains approximately constant during the measurement. Therefore, when EDGE starts in the adsorption-sensitive regime, it remains adsorption-sensitive as the interface becomes progressively covered.

Convected drops and EDGE consequently share an important advantage over classical pendant drops: their mass-transfer resistance does not increase continuously with interface age. Their practical implementations, however, impose different experimental limitations.

In the microtensiometer developed by Alvarez et al. \cite{Alvarez.2010micro, alvarez2012using}, the flow is imposed around an interface attached to the end of a capillary. Increasing the liquid velocity reduces the thickness of the convective boundary layer and accelerates surfactant transport. However, the maximum accessible velocity is limited by the requirement that the interface remains attached, stable, and sufficiently weakly deformed. Increasing the velocity beyond this range risks displacing or detaching the interface. This limitation restricts the extent to which transport can be accelerated, and the reported dynamics for the studied nonionic surfactants remained controlled by convection--diffusion.

In the approach developed by Baret and coworkers \cite{Brosseau2014, Riechers.2016}, the droplet is transported with the flow through a delay line containing successive constrictions and expansions. Its age is determined from its travel time, while the interfacial tension is inferred from its deformation in the expansion regions. This method can access short interface ages, but the interfacial tension measurement is indirect because it relies on a calibrated hydrodynamic relation between droplet shape and interfacial tension. Moreover, the droplet is successively compressed and dilated as it passes through constrictions and expansions. Its interfacial area and shape therefore change during the measurement. The tension is determined only at selected positions along the device, rather than continuously, which complicates the reconstruction of a complete and weakly perturbed tension--time curve.

EDGE avoids the limitations of convected droplets discussed above by separating the renewal flow from the motion of the measured interface. The aqueous solution flows alongside the pores, whereas the oil meniscus remains pinned inside the shallow cavity during most of the adsorption period. Its area and curvature therefore vary only weakly. Increasing the aqueous velocity $U$ reduces the convective depletion-layer thickness according to $\delta_U \propto U^{-1/2}$, and consequently decreases the effective transport length $L_D$. Within the limits imposed by pressure losses and meniscus stability, the renewal velocity can therefore be increased to reduce $L_D$ and move the experiment toward adsorption control without forcing the measured interface to travel through the device.

The EDGE geometry also provides a more direct measurement of interfacial tension. The applied pressure selects a capillary threshold, while the waiting time before droplet release defines the corresponding interface age. The interfacial tension is therefore obtained from the Young--Laplace relation without reconstructing it from the transient deformation of a moving droplet. EDGE thus combines a fixed micrometer-scale transport distance, tunable convective renewal, a nearly stationary interface, and a direct capillary measurement of dynamic interfacial tension.

\section{Sequential identification of interfacial thermodynamics,
transport, and adsorption kinetics}

The interfacial tension measurements contain information about three distinct physical contributions: the thermodynamics of the adsorbed layer, surfactant transport through the liquid, and molecular transfer at the interface. These contributions cannot be determined reliably by fitting all parameters simultaneously to a single dynamic curve. We therefore use a sequential strategy in which the equilibrium, transport, and kinetic parameters are determined from complementary measurements before being combined in a common nonequilibrium model.

A central difficulty, that is also encountered in most studies in the literature \cite{kovalchuk2023surfactant, Kovalchuk.2023}, is that interfacial tension is measured directly, whereas the surface coverage $\phi$ and the subsurface concentration $c_s$ are not. Converting a measured tension--time curve into an adsorption trajectory therefore requires an interfacial thermodynamic model. In particular, the equilibrium equation of state cannot in general be applied at each instant of a dynamic experiment, because this would implicitly assume that the adsorbed layer remains in local equilibrium with the subsurface solution. This assumption is inappropriate when molecular adsorption is the rate-limiting process. We therefore use the Diamant--Andelman \cite{diamant1996kinetics, diamant2001} framework, which expresses the interfacial tension as a function of both $c_s$ and $\phi$.

\subsection{Step 1: equilibrium interfacial thermodynamics}

Equilibrium interfacial-tension measurements are first used to determine the adsorption isotherm and the equation of state. At equilibrium, $c_s=c_b$ and $\phi=\phi_{\mathrm{eq}}$.

For a nonionic surfactant, the equilibrium coverage and interfacial tension are described by the following equations: 

\begin{align}
\frac{\phi_{\mathrm{eq}}}{1-\phi_{\mathrm{eq}}}
& = Kc_b\exp\left(\beta^\star\phi_{\mathrm{eq}}\right)
\label{eq:nonionic_equilibrium_isotherm}
\\ \gamma_{\mathrm{eq}}-\gamma_0
& = RT\Gamma_\infty \left[ \ln(1-\phi_{\mathrm{eq}})
+ \frac{\beta^\star}{2}\phi_{\mathrm{eq}}^2 \right]
\label{eq:nonionic_equilibrium_tension}
\end{align}

where $K$ is the equilibrium adsorption constant, $\Gamma_\infty$ is the maximum surface excess, and $\beta^\star$ accounts for lateral interactions within the adsorbed layer. The Langmuir limit is recovered when $\beta^\star=0$. Fitting $\gamma_{\mathrm{eq}}(c_b)$ therefore determines $K$, $\Gamma_\infty$, and, when required, $\beta^\star$.

Outside equilibrium, the interfacial tension is calculated from the complete Diamant--Andelman \cite{diamant1996kinetics, diamant2001} expression given by the following:

\begin{equation}
\begin{split}
\gamma(c_s,\phi)-\gamma_0 = RT\Gamma_\infty \bigg[
& \phi\ln\phi + (1-\phi)\ln(1-\phi)
\\
&-\phi\ln\left(N_Aa^3c_s\right) -\alpha^\star\phi -\frac{\beta^\star}{2}\phi^2 \bigg]
\end{split}
\label{eq:nonionic_nonequilibrium_tension}
\end{equation}

where $a$ is the molecular length scale and $\alpha^\star$ represents the affinity of the surfactant for the interface. This relation provides the required conversion between the calculated coverage and the measured dynamic interfacial tension without imposing local equilibrium.

For an ionic surfactant, adsorption is additionally opposed by the electrostatic cost of charging the interface. The equilibrium isotherm is written as follows \cite{levine1963discrete, bonfillon1994dynamic, Prosser.2001.review}:

\begin{equation}
\frac{\phi_{\mathrm{eq}}}{1-\phi_{\mathrm{eq}}}
= Kc_b \exp\left[ \beta^\star\phi_{\mathrm{eq}}
- \chi_{\mathrm{el}} \frac{F|\psi_0|}{RT} \right]
\label{eq:ionic_equilibrium_isotherm}
\end{equation}

where $\psi_0$ is the self-consistent surface potential and $\chi_{\mathrm{el}}$ weights the electrostatic contribution to the adsorption free energy. The ionic interfacial tension is calculated using:

\begin{equation}
\gamma(c_s,\phi) = \gamma_{\mathrm{chem}}(c_s,\phi)
+ \gamma_{\mathrm{el}}(c_s,\phi)
\label{eq:ionic_tension_decomposition}
\end{equation}

where $\gamma_{\mathrm{chem}}$ is given by the nonionic
Diamant--Andelman expression (\autoref{eq:nonionic_nonequilibrium_tension}) and $\gamma_{\mathrm{el}}$ accounts for the diffuse-layer free energy. The detailed expressions for $\psi_0$ and $\gamma_{\mathrm{el}}$ are given in the Supplementary Materials. Equilibrium measurements are used to determine the thermodynamic and electrostatic parameters before any dynamic curve is analyzed.

\subsection{Step 2: independent determination of bulk diffusion}

The diffusion coefficient $D$ is determined from dilute classical rising-drop measurements. Under these conditions, the large transport distance and the growth of the diffusion layer make the tension relaxation predominantly diffusion-controlled. The curves are analyzed with the Ward--Tordai model \cite{WARD.1944, WARD.1946}, and the calculus from Nele et al. \cite{nele2022analytically}, using the previously determined equilibrium isotherm (\autoref{eq:nonionic_equilibrium_isotherm}) and equation of state (\autoref{eq:nonionic_equilibrium_tension}).
All interfacial thermodynamic parameters are therefore fixed, leaving $D$ as the only transport parameter determined from the dilute drop data.

\subsection{Step 3: identification of adsorption kinetics with EDGE}
\label{sec:EDGE_kinetic_identification}

The EDGE measurements are used to determine the molecular adsorption
kinetics after the equilibrium thermodynamic parameters and the diffusion coefficient have been determined independently. 

\subsubsection{Nonionic surfactant: adsorption-controlled regime}

In the adsorption-controlled limit, transport from the bulk is sufficiently fast to maintain the subsurface concentration close to the bulk concentration so $c_s \simeq c_b$.
Therefore, no significant depletion develops next to the interface. For a nonionic surfactant, the surface coverage evolves according to

\begin{equation}
\frac{\mathrm{d}\phi}{\mathrm{d}t} =
R_{\mathrm{ads}}^{\mathrm{N}}(c_b,\phi)
= k_a c_b(1-\phi)-k_d\phi
\label{eq:nonionic_adsorption_controlled}
\end{equation}

With $k_d={k_a}/{K}$ and the equilibrium constant $K$ already determined from the equilibrium measurements, $k_a$ is the only kinetic parameter adjusted from the EDGE curves, while $k_d$ defined by the equilibrium constraint.

The interface is assumed to be initially clean, $\phi(0)=0$.
A dedicated calculation presented in the Supporting Information (S3 and Fig.S2) confirms that the residual initial coverage is negligible under the experimental conditions considered.
At each time, the calculated coverage is converted into interfacial tension using the full nonequilibrium Diamant--Andelman expression $\gamma(c_b,\phi)$ given by \autoref{eq:nonionic_nonequilibrium_tension} rather than the equilibrium equation of state. This distinction is essential because, in an adsorption-controlled regime, the adsorbed layer is not assumed to remain at equilibrium with the adjacent solution during relaxation.

The fitted value of $k_a$ is retained as an intrinsic adsorption constant only if the same value describes several concentrations and if transport remains faster than adsorption throughout the measured relaxation. This condition is checked using the regime criterion introduced above, which must remain in the adsorption-sensitive domain, $\Pi<1$. If this condition is not satisfied, subsurface depletion must be explicitly accounted for using the mixed adsorption--transport model described in the Supporting Information S5.

\subsubsection{Ionic surfactant: electrostatic adsorption model}

For an ionic surfactant, we follow the same sequential procedure.
The transport criterion of $\Pi$ will be checked first, and the mixed adsorption--diffusion model is used only when the transport criterion shows that $c_s\simeq c_b$ is not valid.
We first use the electrostatic adsorption law derived from the
Diamant--Andelman \cite{diamant1996kinetics, diamant2001} description. The interfacial free-energy contribution is the following:

\begin{equation}
u_{\mathrm{DA}}(c_s,\phi) =
\beta^\star\phi - \chi_{\mathrm{el}} \frac{F|\psi_0(c_s,\phi)|}{RT}
\label{eq:DA_ionic_energy}
\end{equation}

where $\beta^\star\phi$ accounts for lateral interactions and the second term represents the electrostatic cost associated with charging the interface. The corresponding adsorption rate can be written as:

\begin{equation}
\begin{split}
R_{\mathrm{ads}}^{\mathrm{I}}(c_s,\phi) ={}&
k_a c_s(1-\phi) \exp\left[p\,u_{\mathrm{DA}}(c_s,\phi)\right]
\\ & - k_d\phi\exp\left[-(1-p)u_{\mathrm{DA}}(c_s,\phi)\right]
\end{split}
\label{eq:ionic_ads_rate}
\end{equation}

The coefficient $p$ specifies how the interfacial free-energy difference is distributed between the adsorption and desorption pathways. A symmetric partition, $p=1/2$, is used in the following. In the adsorption-controlled limit, the subsurface concentration is set to $c_s=c_b$ and the surface coverage is directly obtained from: 

\begin{equation}
\frac{\mathrm{d}\phi}{\mathrm{d}t}
=  R_{\mathrm{ads}}^{\mathrm{I}}(c_b,\phi)
\label{eq:ionic_adsorption_controlled_final}
\end{equation}

The calculated coverage is converted into interfacial tension using the complete nonequilibrium ionic equation of state \autoref{eq:ionic_tension_decomposition}, including both the chemical and electrostatic contributions.
The equilibrium, electrostatic, and transport parameters remain fixed in both the adsorption-controlled and mixed analyses. For more details on the calculations, refer to Supporting Information S4. 

\section{Results and discussion}
\label{sec:results}
\subsection{Microfluidic EDGE vs classical rising-drop tensiometry: experimental results} \label{sec:dynamics}

\begin{figure}[ht]
    \centering
    \includegraphics[width=1\linewidth]{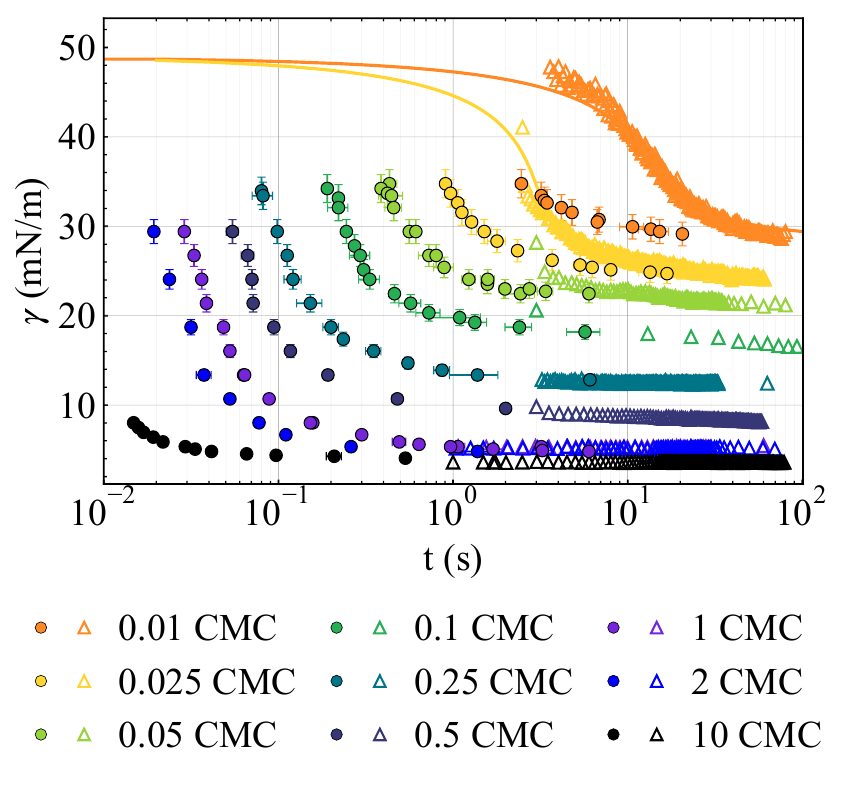}
    \caption{Dynamic interfacial tension of C$_{10}$E$_8$ solutions at the water/hexadecane interface, measured using the EDGE tensiometer (circles) and the rising drop tensiometry (ADT) (triangles). The concentration of the C$_{10}$E$_8$ solutions ranges from $0.01$ CMC to $10$ CMC, with CMC=1.4 \si{\mol \per \cubic \meter}. The error bars of the rising drop tensiometry measurements are included within the triangles. 
    The orange and yellow lines represent the diffusion model from Ward--Tordai \cite{WARD.1944, WARD.1946} using Nele's calculus \cite{nele2022analytically} for D=\num{1.0e-9} \si{\square \meter \per \second} , for c=0.01 CMC and c=0.025 CMC respectively.}
    \label{fig:C10E8_1}
\end{figure}

\autoref{fig:C10E8_1} compares the dynamic interfacial tension of C$_{10}$E$_8$ solutions measured with the EDGE tensiometer and with classical rising-drop tensiometry. The two techniques probe the same water/hexadecane interface and the same concentration range, from $0.01\,\mathrm{CMC}$ to $10\,\mathrm{CMC}$, but they access very different regimes.

A first striking feature is the much faster relaxation measured in the EDGE device. For a given concentration, the decrease of $\gamma$ occurs at significantly shorter times than in the rising-drop experiment. This difference is particularly clear at intermediate and high concentrations, where the EDGE curves reach their plateau within a few tens to hundreds of milliseconds, whereas the rising-drop data evolve over seconds to tens of seconds. The effect is not a change of equilibrium adsorption: at long times, both techniques tend towards comparable limiting surface tensions. Instead, it is a consequence of the different transport conditions. At the lowest concentrations, the diffusion coefficient was extracted using the diffusion-limited Ward–Tordai model \cite{WARD.1944, WARD.1946} applied to the rising-drop data. The best-fit gave D=\num{1.0e-9} \si{\square \meter \per \second}, a value used in the analysis that follows. 

\begin{figure}[ht]
    \centering
    \includegraphics[width=1\linewidth]{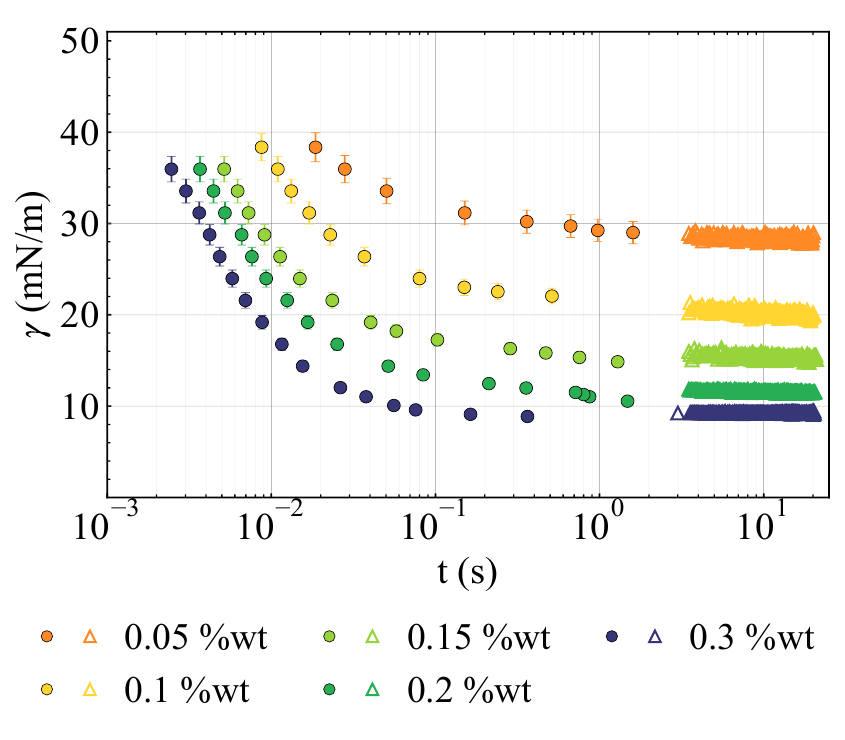}
    \caption{Dynamic interfacial tension of SDS solutions at the water/hexadecane interface, measured using the microfluidic EDGE tensiometer (circles) and the rising drop tensiometer (triangles). The concentration of the surfactant solutions in SDS ranges from 0.05\% wt (= 0.2 CMC) to 0.3\% wt (= 1.2 CMC). Note that CMC$\mathrm{_{SDS}}$=0.25\% wt = 8.7 \si{\mol \per \cubic \meter}.}
    \label{fig:SDS_1}
\end{figure}

\autoref{fig:SDS_1} reports the same comparison for SDS solutions at the water/hexadecane interface. In this case, the contrast between the two techniques is even more pronounced. The EDGE measurements display a rapid decrease of the interfacial tension over short times, with a concentration-dependent relaxation occurring between a few milliseconds and about one second. Increasing the SDS concentration accelerates the decrease of $\gamma$ and lowers the final measured value, as expected from the larger surfactant supply to the interface.

By contrast, the rising-drop measurements show almost horizontal curves over the accessible time window: the classical tensiometer fails to resolve the early adsorption dynamics and essentially only captures the equilibrium interfacial tension. The dynamic regime is therefore accessible only with the EDGE device, which probes much shorter times.
This comparison shows that the dynamic surface tension is not an intrinsic function of time for a given surfactant solution: it depends on how surfactant molecules are transported to the interface — from a nearly quiescent, macroscopic reservoir in the rising-drop tensiometer, to a micrometer-scale reservoir continuously renewed by convection in the EDGE device.

The two techniques therefore probe distinct transport regimes: diffusion-controlled mass transfer over macroscopic distances for the rising drop, versus adsorption under rapid convective renewal and micrometer-scale depletion for EDGE. This is why the same C$_{10}$E$_8$ solution gives different apparent dynamic curves in the two instruments, especially at low concentration, even though the underlying equilibrium adsorption isotherm is identical.

In the following, we use the modeling framework developed above  to analyze the experimental data and to extract both the thermodynamic parameters of the adsorption isotherm and the kinetic parameters governing the mass transport of both studied surfactants. 

\subsection{Model analysis in the C$_{10}$E$_8$ situation}

\subsubsection{Sequential analysis of nonionic C$_{10}$E$_8$ adsorption}
\label{sec:C10E8_analysis}

The C$_{10}$E$_8$ data were analyzed sequentially to separate interfacial thermodynamics, bulk diffusion, and molecular adsorption. The equilibrium measurements were first used to determine the adsorption isotherm and the interfacial equation of state. The diffusion coefficient was obtained independently from classical rising-drop measurements. Finally, the short-time EDGE curves were used to determine the adsorption and desorption rate constants, and the validity of the adsorption-controlled interpretation was verified \emph{a posteriori}.

\subsubsection{Equilibrium interfacial thermodynamics}
\label{sec:C10E8_equilibrium}

The equilibrium interfacial tension was determined from the long-time plateau measured at each C$_{10}$E$_8$ concentration. The equilibrium surface coverage was described by the Langmuir isotherm, and the corresponding interfacial tension was calculated from the equilibrium equation of state, given by \autoref{eq:nonionic_equilibrium_isotherm} and \autoref{eq:nonionic_equilibrium_tension} respectively. These equations are simplified by assuming an ideal adsorbed layer with $\beta^\ast$=0, and give the following:

\begin{equation}
\begin{split}
\phi_{\mathrm{eq}} = \frac{Kc_b}{1+Kc_b} \\
\gamma_{\mathrm{eq}}-\gamma_0 = RT\Gamma_\infty\ln(1- \phi_{\mathrm{eq}})
\end{split}
\label{eq:C10E8_Langmuir}
\end{equation}

\begin{figure}[ht]
    \centering
    \includegraphics[width=1\linewidth]{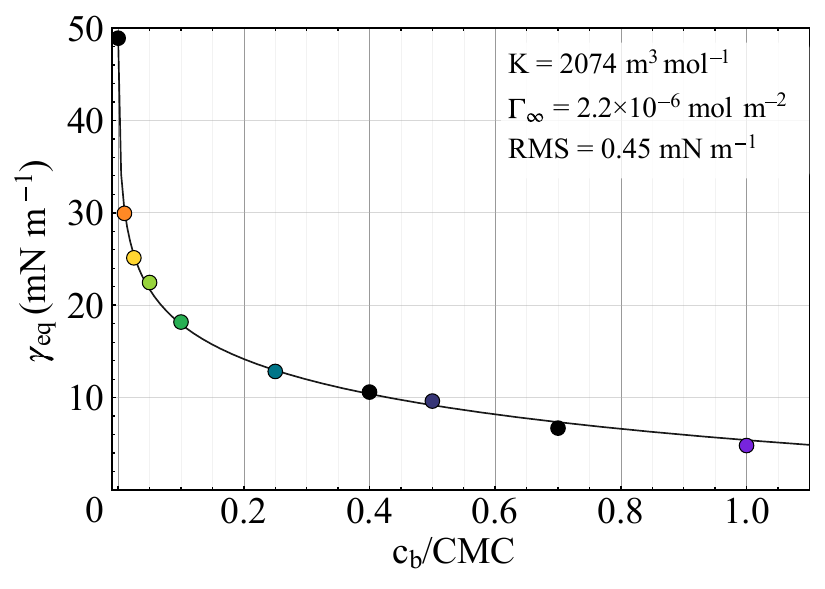}
    \caption{Equilibrium interfacial tension obtained from the long-time plateau of the interfacial tension at the water/hexadecane interface with the best-fit from the classical Langmuir model (line). The colored circles refer to the data obtained with the EDGE tensiometer at long time, while the black circles are data points measured with the classical rising drop tensiometer. Equilibrium was reached at $t=50 \, \mathrm{s}$.}
    \label{fig:C10E8_eq}
\end{figure}

The equilibrium data were well described by this model, as shown in \autoref{fig:C10E8_eq}. The best-fit gave: $\Gamma_\infty$ = \num{2.20e-6} \si{\mol \per \square \meter} and K = \num{2074} \si{\cubic \meter \per \mol}.
These values correspond to a molecular length 
$a=8.70$ \si{\angstrom} and an adsorption affinity
$\alpha/(k_{\mathrm B}T)=15.47$. The RMS deviation between the equilibrium measurements and the model was \num{0.45} \si{\mN \per \meter}. These thermodynamic parameters were subsequently fixed in all dynamic analyses.

\subsubsection{Diffusion coefficient from classical drop tensiometry}
\label{sec:C10E8_diffusion}

The diffusion coefficient of C$_{10}$E$_8$ was determined independently from the dilute rising-drop measurements. In this geometry, the diffusion distance increases continuously with time and the measured relaxation is primarily controlled by transport from the bulk solution.

The dynamic curves were fitted with the Ward--Tordai model \cite{WARD.1944, WARD.1946}, assuming local equilibrium between the adsorbed layer and the subsurface concentration.
The equilibrium parameters $K$, $\Gamma_\infty$, and $\beta^\star=0$ were fixed to the values obtained above, leaving $D$ as the only adjustable transport parameter. The best fit using Nele's calculus \cite{nele2022analytically} gave D=\num{1.0e-9} \si{\square \meter \per \second}, reasonable value given the size of the molecule.
The agreement between the model and the dilute drop measurements (solid lines in \autoref{fig:C10E8_1}) confirms that the classical-drop relaxation can be used to determine the diffusion coefficient independently of the adsorption kinetics. This value of D was therefore fixed in the subsequent EDGE analysis.

\subsubsection{Adsorption kinetics from EDGE measurements}
\label{sec:C10E8_EDGE}

The EDGE measurements were then analyzed using the thermodynamic parameters and diffusion coefficient determined independently above. The effective EDGE transport length was calculated using \autoref{eq:EDGE_transport_length} with the fixed value of D, from the device geometry and under the hydrodynamic renewal conditions, giving $\mathrm{L_D}=7.3$ \si{\um}.

\begin{figure*}[ht]
    \centering
    \includegraphics[width=18cm]{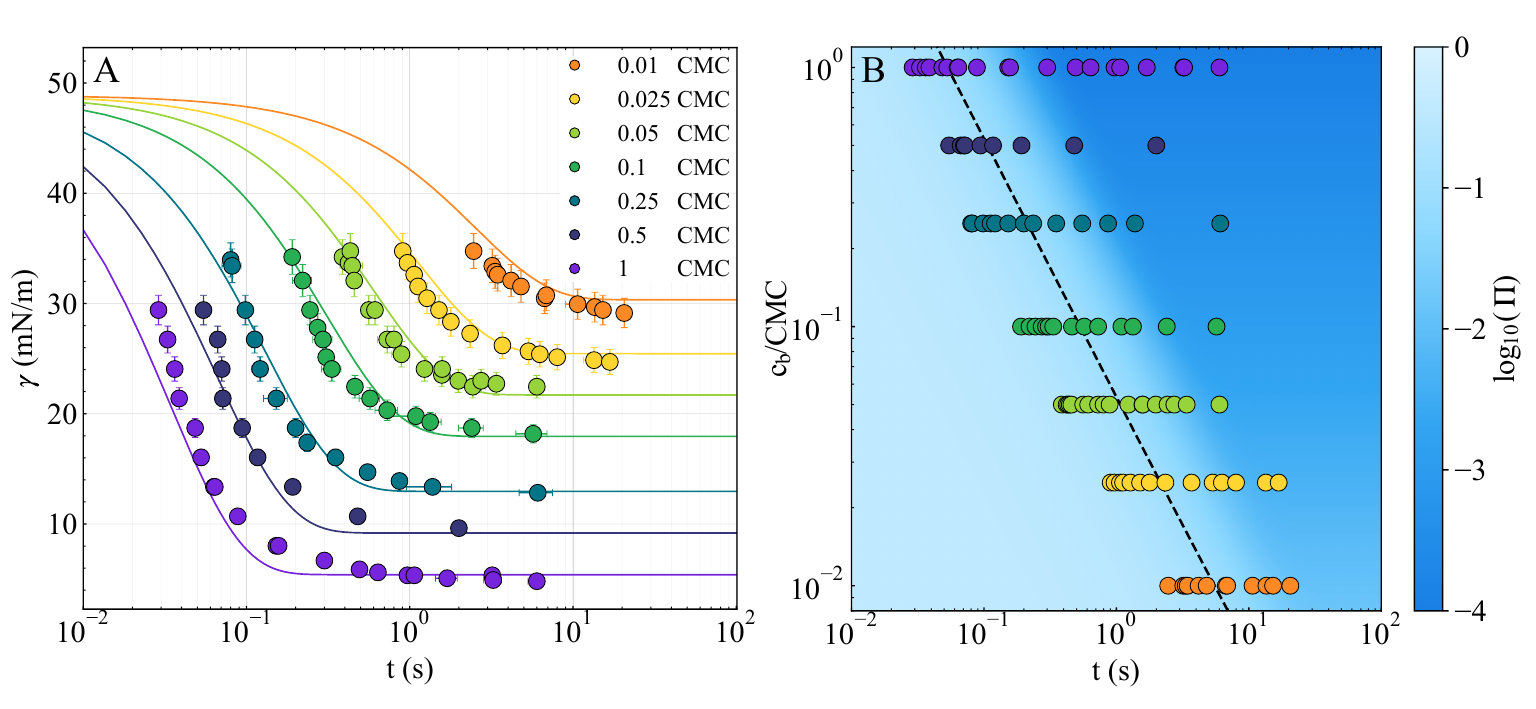}
    \caption{
\textbf{Adsorption-controlled dynamics of C$_{10}$E$_8$ in the EDGE geometry.}
\textbf{(A)} Dynamic interfacial tension measured for C$_{10}$E$_8$ concentrations ranging from 0.01 to 1 CMC (symbols). Solid lines are simultaneous fits of the adsorption-limited model, yielding $k_a$= \num{17.2} \si{\cubic \meter \per \mol \per \second} and $k_d$ = \num{8.3e-3} \si{\per \second}, with an overall RMS error of \num{1.75} \si{\mN \per \meter}. 
\textbf{(B)} Regime map showing $\log_{10}\Pi$ as a function of interface age and normalized bulk concentration $c_b/{\mathrm{CMC}}$, where $\Pi=\Gamma_\infty k_a(1-\phi)/k_m$ compares the characteristic adsorption demand with mass transport toward the interface. Symbols indicate the experimental conditions used in the fit. The dashed line corresponds to $\Pi=0.1$; $\Pi<1$ (and $\mathrm{log(\Pi)<0}$) identifies conditions for which interfacial adsorption kinetics, rather than bulk transport, controls the measured tension dynamics.
}
\label{fig:C10E8_ads}
\end{figure*}

We first analyzed the EDGE curves of C$_{10}$E$_8$ using \autoref{eq:nonionic_adsorption_controlled} with an initially clean interface, $\phi(0)=0$. Because $K$ was fixed by
the equilibrium analysis, $k_a$ was the only dynamically adjusted
parameter. At each time, the calculated coverage was converted into interfacial tension using \autoref{eq:nonionic_nonequilibrium_tension}.

A single pair of kinetic constants described the EDGE curves over the investigated concentration range, gave by the best-fit: $k_a = 17.2$ \si{\cubic \meter \per \mol \per \second} and $
k_d =8.3\times10^{-3}$ \ \si{\per \second}, with the corresponding RMS deviation of \num{1.75} \si{\mN \per \meter}. \autoref{fig:C10E8_ads}A shows the good agreement between the adsorption-limited model and the experimental dynamic interfacial tension data measured by the EDGE device.  
The adsorption-controlled assumption was verified \emph{a posteriori} using the fitted coverage trajectory. A dedicated numerical routine was used to calculate, at every experimental time and for each bulk concentration, the regime parameter $\Pi(t)$ as follows:

\begin{equation}
\Pi(t) = \frac{ \Gamma_\infty k_a(1-\phi(t))} {k_m}
\label{eq:C10E8_Pi}
\end{equation}

with $k_m=(A_f/A_i)D/L_D$. As shown in \autoref{fig:C10E8_ads}B, the calculated value of $\Pi(t)$ remained below unity throughout the analyzed relaxation and for all concentrations included in the fit. Bulk transport was therefore sufficiently fast for the EDGE response to remain sensitive to the intrinsic molecular adsorption rate. This \emph{a posteriori} verification supports the identification of $k_a$ and $k_d$ as intrinsic kinetic constants rather than apparent parameters compensating for an unknown transport limitation. Note that the analysis was also performed using a mixed model, but no significant improvement or change in the kinetic parameters was observed, as detailed in the Supplementary Materials.
The thermodynamic and kinetic parameters obtained for C$_{10}$E$_8$ are summarized in \autoref{tableC}.  Importantly, even though numerous parameters are reported in this table, $k_a$ is the only parameter fitted to the EDGE dynamics; all thermodynamic and transport quantities are independently constrained. This minimizes parameter compensation and allows the extracted adsorption rate to be interpreted as an intrinsic kinetic constant rather than an effective fit parameter. This is  a key strength of our method. \\

\begin{table}[htbp]
\centering
\caption{Thermodynamic, transport, and kinetic parameters obtained for 
C$_{10}$E$_8$ at the water/hexadecane interface.}
\label{tab:C10E8_parameters}
\begin{adjustbox}{max width=\columnwidth}
\begin{tabular}{llll}
\hline
Parameter & Symbol & Value\\
\hline

Maximum surface excess
& $\Gamma_{\infty}$
& \num{2.20e-6} \si{\mol \per \square \meter}\\

Adsorption equilibrium constant
& $K$
& \num{2.07e3} \si{\cubic \meter \per \mol}\\

Lateral interaction parameter
& $\beta^{*}$
& 0 \\

Molecular length
& $a$
& \num{8.70} \si{\angstrom}\\

Adsorption affinity
& $\alpha/(k_{\mathrm B}T)$
& 15.47 \\

Equilibrium-fit coefficient
& $R^{2}$
& 0.998 \\

Equilibrium-fit error
& RMSE$_{\mathrm{eq}}$
& \num{0.45} \si{\mN \per \meter} \\

Diffusion coefficient
& $D$
& \num{1.0e-9} \si{\square \meter \per \second} \\

EDGE transport length
& $\mathrm{L_D}$
& \num{7.3} \si{\um} \\

Mass-transfer coefficient
& $k_m$
& \num{1.89e-4} \si{\meter \per \second}\\

Adsorption rate constant
& $k_a$
& \num{17.1} \si{\cubic \meter \per \mol \per \second} \\

Desorption rate constant
& $k_d$
& \num{8.3e-3} \si{\per \second}\\

Dynamic-fit error
& RMSE$_{\mathrm{dyn}}$
& \num{1.75} \si{\mN \per \meter}\\

\hline
\end{tabular}
\end{adjustbox}
\label{tableC}
\end{table}

\subsection{Model analysis for SDS}

Unlike C$_{10}$E$_8$, SDS is an ionic surfactant. Its adsorption therefore involves not only transport through the liquid and molecular transfer to the interface, but also the electrostatic cost associated with charging the adsorbed layer. The SDS analysis consequently requires an ionic thermodynamic and kinetic description.
A second important difference is experimental. For C$_{10}$E$_8$, the rising-drop measurements provide a sufficiently extended diffusion-controlled relaxation to determine the diffusion coefficient independently before analyzing the EDGE data. This is not possible for SDS. Over the accessible time window, the rising-drop curves are already close to their long-time values and do not resolve the early adsorption dynamics. They are therefore used primarily to constrain equilibrium interfacial thermodynamics, while the SDS diffusion coefficient is fixed
from literature data. We use D = \num{5.0e-10} \si{\square \meter \per \second}, as reported for SDS at a closely related water--alkane interface \cite{javadi2010effects}. With the EDGE geometry and renewal conditions, the effective transport length is $\mathrm{L_D}=5.77$ \si{\um} and $k_m=$ \num{6.67e-5} \si{\meter \per \second}.
Thus, as for C$_{10}$E$_8$, the transport parameters entering the EDGE analysis are fixed independently of the dynamic fits. The analysis can then focus on the additional thermodynamic and kinetic effects associated with the ionic nature of SDS.

\subsubsection{Equilibrium thermodynamics of SDS}

We first determine the ionic interfacial thermodynamics independently of the dynamic EDGE measurements. For an ionic surfactant, adsorption is opposed by the electrostatic work required to charge the interface.
The equilibrium isotherm and the corresponding interfacial tension are given by \autoref{eq:ionic_equilibrium_isotherm} and \autoref{eq:ionic_tension_decomposition} respectively. The details of the calculations are given in the Supporting Information S4.

\begin{figure}[ht]
    \centering
    \includegraphics[width=\linewidth]{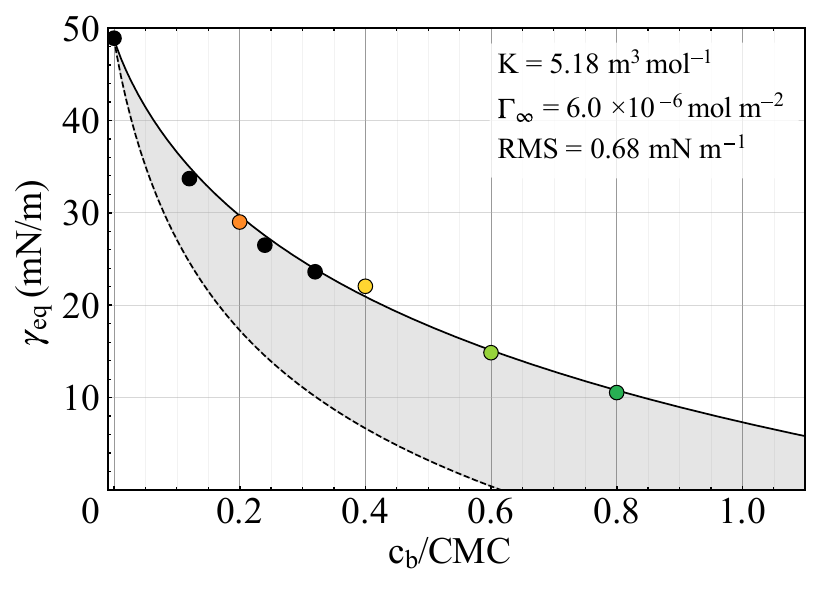}
    \caption{
    Equilibrium interfacial tension of SDS at the water--hexadecane interface as a function of normalized bulk concentration $c_b/{\mathrm{CMC}}$.
    Colored circles correspond to the equilibrium values associated with the EDGE concentrations, and the black circle at $c_b=0$ corresponds to the clean interface,
    $\gamma_0=$ \num{48.9} \si{\mN \per \meter }.
    The solid black line is the total equilibrium fit obtained with the ionic adsorption model and $\beta^\star=0$.
    The dashed line represents the chemical contribution $\gamma_{\mathrm{chem}}$. The shaded region represents $\gamma_{\mathrm{el}}=\gamma_{\mathrm{eq}} -\gamma_{\mathrm{chem}}$, i.e. the positive free-energy contribution associated with charging the interface and forming the diffuse ionic layer.
    }
    \label{fig:sds_eq}
\end{figure}

The ionic equilibrium model contains several thermodynamic and electrostatic parameters that may be strongly correlated if they are all adjusted simultaneously. We therefore begin with the simplest description compatible with the data. Lateral interactions are neglected by fixing $\beta^\star=0$, so that deviations from Langmuir adsorption arise here from the electrostatic contribution rather than from an additional adjustable Frumkin interaction.

We also fix the electrostatic adsorption factor to $\chi_{\mathrm{el}}=0.14$.
This value is physically consistent with the strong counterion association known for SDS.
Indeed, if $\alpha$ denotes the effective ionization fraction, the interfacial charge density is reduced according to
$\sigma_{\mathrm{eff}}=\alpha F\Gamma$ and, because the Gouy--Chapman surface potential depends nonlinearly on the surface charge,
$\chi_{\mathrm{el}}$  can be approximated as
$\chi_{\mathrm{el}}\simeq\alpha|\psi_{\mathrm{GC}}(\alpha\sigma)|/|\psi_{\mathrm{GC}}(\sigma)|$.
Using $\alpha\simeq0.20$, consistent with reported counterion dissociation fractions for SDS aggregates \cite{Benrraou2003},
gives $\chi_{\mathrm{el}}\simeq0.14$ over the present concentration range. We therefore use $\chi_{\mathrm{el}}$ as an effective weighting of the electrostatic contribution to the adsorption free energy, rather than as a direct measurement of the ionization degree of the planar interface.

The interfacial tension contains a separate diffuse-layer contribution, $\gamma_{\mathrm{el}}$. Its magnitude is scaled by the prefactor $\chi_{\gamma}$, which enters the interfacial equation of state but not the adsorption free-energy barrier. Thus, $\chi_{\mathrm{el}}$ and $\chi_{\gamma}$ play distinct roles in the model: the former controls the electrostatic penalty for adsorption, whereas the latter sets the magnitude of the diffuse-layer contribution to the measured
interfacial tension.

With $\beta^\star$ and $\chi_{\mathrm{el}}$ fixed, the equilibrium fit determines $\Gamma_\infty$, $K$, and $\chi_{\gamma}$. The best fit gives: $\Gamma_\infty=$ \num{6.0e-6} \si{\mol \per \square \meter}, $K=5.18$ \si{\cubic \meter \per \mol} and $\chi_{\gamma}=0.15$.

With the present sign convention, the diffuse-layer free-energy
contribution $\gamma_{\mathrm{el}}$ is positive and therefore
partially compensates the decrease in interfacial tension produced
by adsorption. As shown in \autoref{fig:sds_eq}, this contribution reaches approximately 15 \si{\mN \per \meter} over the investigated concentration range.
All equilibrium parameters are subsequently fixed before analyzing the dynamic EDGE measurements.

\subsubsection{Adsorption-controlled description}

We first test whether the SDS dynamics can be described in the same adsorption-controlled limit identified for C$_{10}$E$_8$. In this limit, transport is assumed to be sufficiently fast to maintain $c_s=c_b$. No additional phenomenological kinetic barrier is introduced. 

\begin{figure}[ht]
    \centering
    \includegraphics[width=1\linewidth]{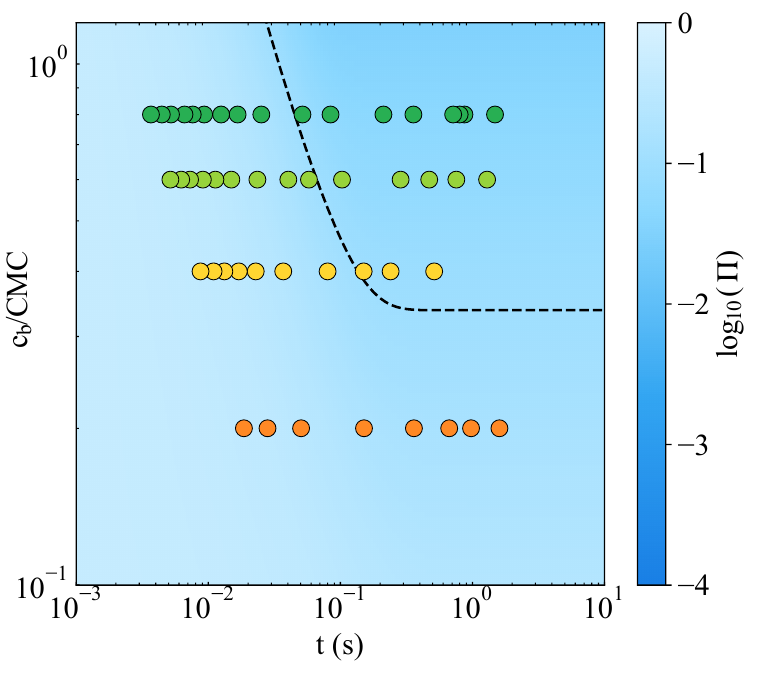}
    \caption{\textbf{Interfacial dynamics of SDS in the EDGE geometry.} 
    Regime map showing $\log_{10}(\Pi_{\mathrm{eff}})$ as a function of interface age and normalized bulk concentration, using the best-fit from the adsorption model (assuming $c_b=c_s$) with $k_a$= \num{9.63} \si{\cubic \meter \per \second}, and $k_d=$ \num{1.86} \si{\per \second} with an overal RMS error of $\mathrm{RMSE}_{\mathrm{ads}}=$ 1.32 \si{\mN \per \meter}. Symbols indicate the experimental conditions included in the fit. $\Pi_{\mathrm{eff}}<1$ identifies conditions for which the EDGE measurements remain sensitive to interfacial adsorption kinetics, whereas $\Pi_{\mathrm{eff}}\ll1$ would correspond to a strictly adsorption-controlled regime. Here, $\Pi_{\mathrm{eff}}$ reaches values of order 0.5, indicating that adsorption remains identifiable but transport effects are not negligible. }
\label{fig:sds_mix}
\end{figure}

With $\beta^\star=0$, the ionic adsorption rate is given by $R^I_{ads}(c_b,\phi)$ from \autoref{eq:ionic_ads_rate}
with $u_\mathrm{DA}= -\chi_{\mathrm{el}}{F|\psi_0|}/{RT}$ and p=1/2.
Because $K$ and all other equilibrium parameters have already been
determined, $k_a$ is the only dynamically adjusted parameter.
A simultaneous fit of the four SDS concentrations gives: 
$k_a = 9.63~\mathrm{m^3\,mol^{-1}\,s^{-1}}$, $k_d=$ 1.86 \si{\per \second}, with $\mathrm{RMSE}_{\mathrm{ads}}=$ 1.32 \si{\mN \per \meter} and $R^2_{\mathrm{ads}}$=0.978. 
The adsorption-only model therefore captures the overall timescale and concentration dependence of the SDS relaxation reasonably well.
However, Figure S3 shows that it does not reproduce the complete shape of the experimental curves: systematic deviations remain during the relaxation. This suggests that the approximation $c_s=c_b$ may not be fully satisfied.
Because $k_m$ is independently known, this assumption can be tested without readjusting the transport rate. For the ionic adsorption law we define:

\begin{equation}
\Pi_{\mathrm{eff}}(t) = \frac{\Gamma_\infty k_a(1-\phi(t))} {k_m} \exp(pu)
\end{equation}

Over all experimental conditions, as shown in \autoref{fig:sds_mix}, $\Pi_{\mathrm{eff,max}} \simeq 0.50$. The criterion remains below unity, confirming that the measurements remain sensitive to molecular adsorption. However, $\Pi_{\mathrm{eff}}$ is not much smaller than unity. Transport is therefore not sufficiently faster than adsorption to guarantee $c_s=c_b$ throughout the complete relaxation.
The systematic shift observed in Figure S3, together with this independent regime criterion, motivate a mixed description in which the same interfacial adsorption kinetics are retained but the subsurface concentration is allowed to evolve in time.

\subsubsection{Mixed adsorption--transport dynamics}
We therefore introduce a finite subsurface reservoir whose concentration can transiently depart from the bulk value. Its dynamics is written as follows:

\begin{equation}
\frac{dc_s}{dt} = \frac{D}{L_D L_s}(c_b-c_s)
- \frac{A_i}{A_f} \frac{\Gamma_\infty}{L_s} R^I_{\mathrm{ads}}(c_s,\phi)
\label{eq:SDS_cs}
\end{equation}

together with

\begin{equation}
\frac{d\phi}{dt} = R^I_{\mathrm{ads}}(c_s,\phi)
\label{eq:SDS_phi}
\end{equation}

Here, $L_s$ is an effective storage length. It should not be interpreted as the thickness of a molecular interfacial layer. Rather, it represents the effective depth of the near-interface liquid reservoir whose concentration can be transiently depleted by adsorption before being replenished from the bulk.

In the absence of adsorption, \autoref{eq:SDS_cs} gives a
characteristic replenishment time:

\begin{equation}
\tau_s =\frac{L_D L_s}{D}
\end{equation}

Importantly, no new interfacial kinetic mechanism is introduced in this second description. The same electrostatic adsorption law is retained for $R^I_{\mathrm{ads}}(c_s,\phi)$ using \autoref{eq:ionic_ads_rate}, with $\beta^\star=0$, $\chi_{\mathrm{el}}=0.14$, and all equilibrium and transport parameters fixed. Only $k_a$ and $L_s$ are adjusted.

\begin{figure*}[!ht]
    \centering
    \includegraphics[width=1\linewidth]{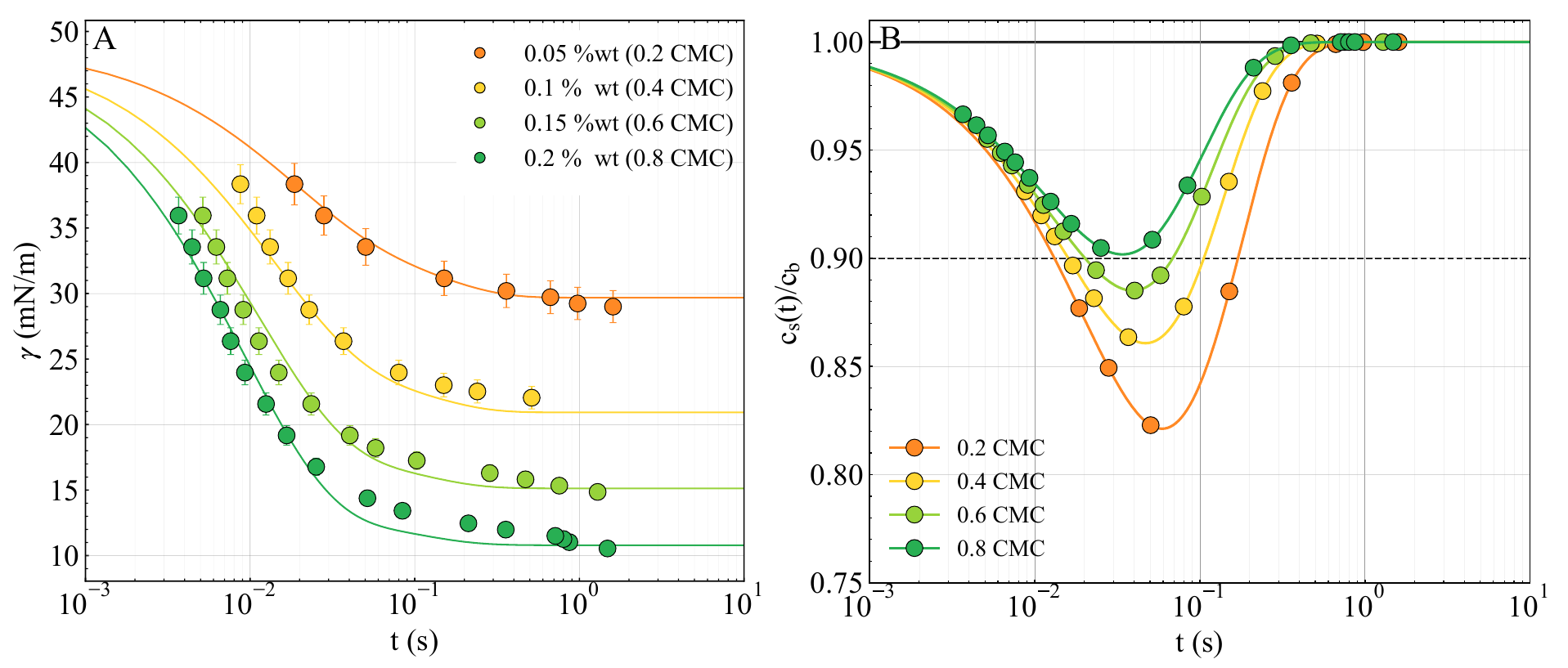}
    \caption{ \textbf{Mixed-adsorption dynamics of SDS in the EDGE geometry.}
    \textbf{(A)} Dynamic interfacial tension measured for SDS concentrations ranging from 0.05\% wt (0.2 CMC) to 0.2\% wt (0.8 CMC) (symbols). Solid lines are simultaneous fits of the mixed model (${c_b} \neq {c_s}$), yielding $k_a$= \num{10.62} \si{\cubic \meter \per \mol \per \second}, $k_d$ = \num{2.05} \si{\per \second} and  $L_s=$ \num{6.95} \si{\um}, with an overall $\mathrm{RMS_{mix}}=$\num{0.936} \si{\mN \per \meter}. 
    \textbf{(B)} Analysis of transient subsurface storage of SDS dynamics in the EDGE geometry. Evolution of the normalized subsurface concentration $c_s/c_b$. $c_s(t)$ is calculated using the mixed-model described above. The transient decrease of $c_s/c_b$ quantifies local depletion caused by adsorption, whereas its subsequent recovery reflects diffusive replenishment from the bulk.
    }
    \label{fig:sds_mixed}
\end{figure*}

The simultaneous fit gives $k_a$= \num{10.62} \si{\cubic \meter \per \second}, $L_s=$ \num{6.95} \si{\um}, and therefore $k_d=$ \num{2.05} \si{\per \second}.
The agreement improves to $\mathrm{RMSE}_{\mathrm{mix}}=$
0.988 \si{\mN \per \meter} and $R^2_{\mathrm{mix}}=0.99$.
\autoref{fig:sds_mixed} shows that introducing a finite subsurface response improves not only the global fit but also the shape of the relaxation. 
The calculated $c_s/c_b$ trajectories in \autoref{fig:sds_mixed}B provide a direct interpretation: rapid adsorption initially depletes the near-interface reservoir, followed by diffusive replenishment on a characteristic timescale of approximately $70~\mathrm{ms}$.\\

The fitted adsorption constant changes only moderately, from $9.63$ to $10.62$ \si{\cubic \meter \per \mol \per \second }, when the transient transport contribution is introduced. This relative stability supports the conclusion that EDGE remains primarily sensitive to the interfacial adsorption kinetics, while the improved description of the curve shape requires accounting for a finite subsurface transport memory.
Thus, for SDS, electrostatic effects are required to describe the ionic adsorption thermodynamics and kinetics, but electrostatics alone do not fully account for the measured relaxation. A transient subsurface depletion and replenishment provides the additional contribution needed to reproduce the dynamics without introducing an extra coverage-dependent phenomenological kinetic barrier. To summarize our results, the thermodynamic, transport, and kinetic parameters obtained for SDS are summarized in \autoref{tab:SDS_parameters}.

\begin{table}[t]
\centering
\caption{Thermodynamic, transport, and kinetic parameters obtained for SDS at the water/hexadecane interface.}
\label{tab:SDS_parameters}
\begin{adjustbox}{max width=\columnwidth}
\begin{tabular}{lll}
\toprule
Quantity & Symbol & Value \\
\midrule

Critical micelle concentration
& ${\mathrm{CMC}}$
& 8.67 \si{\mol \per \cubic \meter} \\

Clean interfacial tension
& $\gamma_0$
& 48.9 \si{\mN \per \meter} \\

Maximum surface excess
& $\Gamma_\infty$
& \num{6.0e-6} \si{\mol \per \square \meter}\\

Equilibrium adsorption constant
& $K$
& 5.18 \si{\cubic \meter \per \mol} \\

Lateral interaction parameter
& $\beta^\star$
& 0 \\

Electrostatic adsorption factor
& $\chi_{\mathrm{el}}$
& 0.14 \\

Electrostatic tension factor
& $\chi_\gamma$
& 0.15 \\

Molecular length
& $a$
& 5.26 \si{\angstrom} \\

Adsorption affinity
& $\alpha/(k_{\mathrm B}T)$
& 10.99 \\

Equilibrium-fit coefficient
& $R^2_{\mathrm{eq}}$
& 0.998 \\

Equilibrium-fit error
& $\mathrm{RMSE}_{\mathrm{eq}}$
& 0.68 \si{\mN \per \meter} \\

Diffusion coefficient
& $D$
& \num{5.0e-10} \si{\square \meter \per \second}\\

EDGE transport length
& $L_D$
& 5.77 \si{\um} \\

Mass-transfer coefficient
& $k_m$
& \num{6.28e-5} \si{\meter \per \second} \\

Adsorption rate, $c_s=c_b$
& $k_a$
& 9.63 \si{\cubic \meter \per \mol}\\

Desorption rate, $c_s=c_b$
& $k_d$
& 1.86 \si{\per \second} \\

Adsorption-only error
& $\mathrm{RMSE}_{\mathrm{ads}}$
& 1.32 \si{\mN \per \meter}  \\

Maximum effective regime parameter
& $\Pi_{\mathrm{eff,max}}$
& 0.50 \\

Mixed-model adsorption rate
& $k_a$
& 10.62 \si{\cubic \meter \per \mol} \\

Subsurface storage length
& $L_s$
& 6.95 \si{\um} \\

Mixed-model desorption rate
& $k_d$
& 2.05 \si{\per \second} \\

Subsurface relaxation time
& $\tau_s=L_D L_s/D$
& \num{6.92e-2} \si{\second} \\

Mixed-model error
& $\mathrm{RMSE}_{\mathrm{mix}}$
& 0.936 \si{\mN\per\meter} \\

Mixed-model coefficient
& $R^2_{\mathrm{mix}}$
& 0.988  \\

\bottomrule
\end{tabular}
\end{adjustbox}
\end{table}

\section{Conclusion}
\label{sec:conclusion}
We have developed a strategy to identify surfactant adsorption kinetics from dynamic interfacial-tension measurements by explicitly accounting for the transport conditions imposed by the experimental geometry. The central result is that access to short interface ages is not, by itself, sufficient to determine intrinsic adsorption kinetics. The measured relaxation reflects interfacial thermodynamics, molecular adsorption, and surfactant transport, and these contributions must be independently constrained before an adsorption rate constant can be identified.
The EDGE geometry provides this separation by combining a nearly stationary interface with micrometer-scale transport distances and controlled hydrodynamic renewal. Equilibrium measurements first determine the adsorption isotherm and interfacial equation of state, while independent transport measurements or estimates constrain the diffusion coefficient and the EDGE mass-transfer resistance. A regime criterion then establishes whether a given dynamic measurement is actually sensitive to molecular adsorption.
For the nonionic surfactant C$_{10}$E$_8$, the EDGE measurements remain in the adsorption-sensitive regime over the investigated conditions. A single adsorption rate constant, $k_a$= \num{17.1} \si{\cubic \meter \per \second} , describes all concentrations once equilibrium thermodynamics and transport are independently fixed. The corresponding kinetic constants can therefore be interpreted as intrinsic interfacial parameters rather than effective quantities compensating for an unknown mass-transfer resistance.
The ionic surfactant SDS illustrates a more complex regime. The electrostatic contribution to the adsorption free energy is required to describe its equilibrium and dynamic behavior, but an adsorption-only model does not reproduce the complete relaxation.
Allowing the near-interface concentration to evolve accounts for a transient depletion followed by diffusive replenishment. The resulting storage length, $L_s =$ 6.95 \si{\um}, corresponds to a characteristic replenishment time of approximately $70~\mathrm{ms}$ and yields $k_a$= \num{10.62} \si{\cubic \meter \per \second}. Importantly, introducing this transport memory changes the inferred adsorption rate only moderately and removes the need for an additional phenomenological coverage-dependent kinetic barrier.
These results establish a general distinction between measuring a young interface and measuring intrinsic adsorption kinetics. The latter requires not only sufficient temporal resolution, but also quantitative control of the transport history experienced by the interface. By combining controlled geometry, independently constrained transport, and nonequilibrium interfacial thermodynamics, EDGE tensiometry provides a route to distinguish molecular adsorption from diffusion and transient subsurface depletion, and to determine when adsorption rate constants can be meaningfully transferred between experimental geometries.

\section{Materials and methods} 
\label{sec:Mat_Met}

\subsection{Materials}
Hexadecane (Reagent Plus 99\%, Sigma-Aldrich) was used as the dispersed phase for all experiments. Hexadecane was first brought into contact with pure water to remove any impurities into the aqueous phase. Two surfactants were used: nonionic 3,6,9,12,15,18,21,24-octaoxatetratriacontan-1-ol (or octaethylene glycol monodecyl ether) C$_{10}$E$_8$, (BioXtra \( \geq 98.0\%\), Sigma-Aldrich); and ionic sodium dodecylsulfate $SDS$ (ACS reagent $\geq99.0\% $, Sigma-Aldrich). Aqueous solutions were prepared with ultrapure water (Milli-Q, Merck Millipore) and used as the continuous phase. For the microfluidic experiments, all aqueous solutions were filtered using 0.22 $\mu m$ (Merck, Germany), or 0.2 $\mu m$ PES filters (VWR) before use. Chip cleaning procedures utilize ethanol (96 \% v/v, VWR International B.V., the Netherlands) and a piranha solution, prepared in a 3:1 volume ratio of sulfuric acid (96\% purity, Sigma-Aldrich, USA) to 35 \% wt hydrogen peroxide (Sigma-Aldrich, USA).

\subsection{Rising drop tensiometry measurements}
\label{sec:ADT}

For long-term measurements, a classical rising drop tensiometer \cite{berry2015measurement} was used. In order to define an initial state of the interface that is as free as possible from surfactant, it is preferable to place the surfactant in solution in the external phase and not in the droplet. Few droplets are always discarded before starting the measurement. As the oil is less dense than the surfactant solutions, we use  a rising drop setup: a droplet of hexadecane rises into the water phase, allowing proper interaction with the surfactant. The accessible time range is limited to short periods by the formation of the drop and to long periods by mechanical vibrations or by an equilibrium tension that is too low, which can cause the drop to fall or rise. To conduct these experiments, we used a commercial setup from Tracker, Teclis, France, called an Automated Drop Tensiometer (ADT). This setup enables us to measure dynamic interfacial tension over a timescale ranging from 1 second to one hour. For all the solutions tested, equilibrium is reached within 50 s.

\subsubsection{Limitations of the method}

The microfluidic EDGE tensiometer is suitable for values of $\gamma$ below $\approx 45$ \si{\mN\per\meter}. Note that with the EDGE tensiometer, it is almost impossible to measure the interfacial tension of pure fluids at very long time. The rising drop tensiometer is best suited for measuring equilibrium interfacial tension $\gamma_{eq}$ while the microfluidic EDGE tensiometer excels at capturing short-time adsorption kinetics. Both techniques handle a wide range of surfactants and concentrations, and together they allow diffusion and adsorption to be resolved independently.

\subsection{Data treatment}
Videos and images were recorded with a high-speed camera (FASTCAM SA-Z, Photron Limited, Japan) connected to the inverted microscope (Axiovert 200 MAT, Carl Zeiss B.V., The Netherlands). In each experiment, three videos were recorded (at a frame rate of 100 000 fps) for the measurement of droplet formation frequency (\(f_0\)) at each pore. \(f_0\) was obtained by counting the number of droplets formed at the pore within a set period of time. The pore frequency \(f_0\) value was first averaged over three independent recordings and then over the eight pores on the plateau. The surfactant mass transport time (i.e., the droplet formation time) is given by \(1/f_0\). Error bars on time are calculated from the standard deviation of droplet counts of three recorded videos. Error bars for the measurement of interfacial tension are calculated using the uncertainty propagation method.  

\section{Authors contribution}
\textbf{C.B.:} Writing - original draft, Visualisation, Validation, Methodology, Data acquisition, Data curation, Formal analysis, Conceptualization. \textbf{B.D.:} Writing - review \& editing, Visualisation, Validation, Methodology, Conceptualization. \textbf{C.D.:} Writing - review \& editing, Supervision, Methodology, Funding acquisition, Conceptualization. \textbf{K.S.:} Writing - review \& editing, Supervision, Methodology, Funding acquisition, Conceptualization. \textbf{A.C.:} Writing - original draft, Writing - review \& editing, Supervision, Methodology, Funding acquisition, Formal analysis, Conceptualization.

\section{Acknowledgements}
This investigation is supported by IFP Energies nouvelles doctorate fundings. The authors thank the Laboratory of Process Engineering of Wageningen University and Research (WUR, The Netherlands) for access to the laboratory and equipments. C.B. warmly thanks Professor Benoit Scheid from the Université Libre de Bruxelles (ULB, Belgium) for very interesting discussions regarding the hydrodynamics in this specific geometry. C.B. acknowledges the support and guidance of Marie Marsiglia. C.B. acknowledges the help of Eric Trant on the optimization of the solvers. 

\newpage 
\addcontentsline{toc}{section}{References}
\bibliographystyle{ieeetr}
\bibliography{References}

@article{diamant1996kinetics,
  title={Kinetics of surfactant adsorption at fluid- fluid interfaces},
  author={Diamant, Haim and Andelman, David},
  journal={The Journal of Physical Chemistry},
  volume={100},
  number={32},
  pages={13732--13742},
  year={1996},
  publisher={ACS Publications}
}

@article{diamant2001,
  author  = {Diamant, Haim and Ariel, Gil and Andelman, David},
  title   = {Kinetics of surfactant adsorption: the free energy approach},
  journal = {Colloids and Surfaces A: Physicochemical and Engineering Aspects},
  volume  = {183--185},
  pages   = {259--276},
  year    = {2001},
  doi     = {10.1016/S0927-7757(01)00553-2}
}

@article{levine1963discrete,
  title={The discrete-ion effect and surface potentials of ionized monolayers},
  author={Levine, S and Mingins, J and Bell, GM},
  journal={The Journal of Physical Chemistry},
  volume={67},
  number={10},
  pages={2095--2105},
  year={1963},
  publisher={ACS Publications}
}

@article{bonfillon1994dynamic,
  title={Dynamic surface tension of ionic surfactant solutions},
  author={Bonfillon, A and Sicoli, F and Langevin, DJJC},
  journal={Journal of colloid and interface science},
  volume={168},
  number={2},
  pages={497--504},
  year={1994},
  publisher={Elsevier}
}

@article{Prosser.2001.review,
 author = {Prosser, Alissa J. and Franses, Elias I.},
 year = {2001},
 title = {Adsorption and surface tension of ionic surfactants at the air--water interface: review and evaluation of equilibrium models},
 url = {https://www.sciencedirect.com/science/article/pii/S0927775700007068},
 pages = {1--40},
 volume = {178},
 number = {1-3},
 issn = {0927-7757},
 journal = {Colloids and Surfaces A: Physicochemical and Engineering Aspects},
 doi = {10.1016/S0927-7757(00)00706-8}
}

@article{cao2026heat,
  title={Heat treatment influences adsorption of potato protein at the oil-water interface and emulsion droplet stability at short timescales},
  author={Cao, Jiarui and Santos, Tatiana Porto and del Prado Castillo, Patricio Cid and Deng, Boxin and Corstens, Meinou and Schro{\"e}n, Karin},
  journal={Journal of Colloid and Interface Science},
  pages={141023},
  year={2026},
  publisher={Elsevier}
}

@article{alvarez2012using,
  title={Using bulk convection in a microtensiometer to approach kinetic-limited surfactant dynamics at fluid--fluid interfaces},
  author={Alvarez, Nicolas J and Vogus, Douglas R and Walker, Lynn M and Anna, Shelley L},
  journal={Journal of colloid and interface science},
  volume={372},
  number={1},
  pages={183--191},
  year={2012},
  publisher={Elsevier}
}

@article{Benrraou2003,
  author  = {Benrraou, Mohamed and Bales, Barney L. and Zana, Raoul},
  title   = {Effect of the Nature of the Counterion on the Properties of Anionic Surfactants. 1. Cmc, Ionization Degree at the Cmc and Aggregation Number of Micelles of Sodium, Cesium, Tetramethylammonium, Tetraethylammonium, Tetrapropylammonium, and Tetrabutylammonium Dodecyl Sulfates},
  journal = {The Journal of Physical Chemistry B},
  year    = {2003},
  volume  = {107},
  number  = {48},
  pages   = {13432--13440},
  doi     = {10.1021/jp021714u}
}

@article{javadi2010effects,
  title={Effects of dodecanol on the adsorption kinetics of SDS at the water--hexane interface},
  author={Javadi, A and Mucic, N and Vollhardt, D and Fainerman, VB and Miller, R},
  journal={Journal of colloid and interface science},
  volume={351},
  number={2},
  pages={537--541},
  year={2010},
  publisher={Elsevier}
}

@article{maan2011spontaneous,
  author  = {Maan, Abid Aslam and Schro{\"e}n, Karin and Boom, Remko M.},
  title   = {Spontaneous droplet formation techniques for monodisperse emulsions preparation: Perspectives for food applications},
  journal = {Journal of Food Engineering},
  year    = {2011},
  volume  = {107},
  number  = {3--4},
  pages   = {334--346},
  doi     = {10.1016/j.jfoodeng.2011.07.008}
}

@article{schultz2004high,
  author  = {Schultz, Stefan and Wagner, Gerhard and Urban, Kai and Ulrich, Joachim},
  title   = {High-Pressure Homogenization as a Process for Emulsion Formation},
  journal = {Chemical Engineering \& Technology},
  year    = {2004},
  volume  = {27},
  number  = {4},
  pages   = {361--368},
  doi     = {10.1002/ceat.200406111}
}

@article{malysa2005influence,
  author  = {Ma{\l}ysa, Kazimierz and Krasowska, Marta and Krzan, Marcel},
  title   = {Influence of surface active substances on bubble motion and collision with various interfaces},
  journal = {Advances in Colloid and Interface Science},
  year    = {2005},
  volume  = {114--115},
  pages   = {205--225},
  doi     = {10.1016/j.cis.2004.08.004}
}

@article{Brosseau2014,
  author  = {Brosseau, Quentin and Vrignon, J{\'e}r{\'e}my and Baret, Jean-Christophe},
  title   = {Microfluidic Dynamic Interfacial Tensiometry ({\ensuremath{\mu}DIT})},
  journal = {Soft Matter},
  year    = {2014},
  volume  = {10},
  number  = {17},
  pages   = {3066--3076},
  doi     = {10.1039/C3SM52543K}
}

@article{Riechers.2016,
  author  = {Riechers, Birte and Maes, Florine and Akoury, Elias and Semin, Benoit and Gruner, Philipp and Baret, Jean-Christophe},
  title   = {Surfactant adsorption kinetics in microfluidics},
  journal = {Proceedings of the National Academy of Sciences of the United States of America},
  year    = {2016},
  volume  = {113},
  number  = {41},
  pages   = {11465--11470},
  doi     = {10.1073/pnas.1604307113}
}

@article{Deng2022EDGE,
  author  = {Deng, Boxin and Schro{\"e}n, Karin and Steegmans, Maartje and de Ruiter, Jolet},
  title   = {Capillary pressure-based measurement of dynamic interfacial tension in a spontaneous microfluidic sensor},
  journal = {Lab on a Chip},
  year    = {2022},
  volume  = {22},
  number  = {20},
  pages   = {3860--3868},
  doi     = {10.1039/D2LC00545J}
}

@article{Santos2024interfacial,
  author  = {Porto Santos, Tatiana and Deng, Boxin and Corstens, Meinou and Berton-Carabin, Claire and Schro{\"e}n, Karin},
  title   = {Interfacial protein adsorption behavior can be connected across a wide range of timescales using the microfluidic {EDGE} ({Edge}-based droplet {GEneration}) tensiometer},
  journal = {Journal of Colloid and Interface Science},
  year    = {2024},
  volume  = {674},
  pages   = {951--958},
  doi     = {10.1016/j.jcis.2024.06.200}
}

@article{christov2006maximum,
  title={Maximum bubble pressure method: Universal surface age and transport mechanisms in surfactant solutions},
  author={Christov, Nikolay C and Danov, Krassimir D and Kralchevsky, Peter A and Ananthapadmanabhan, Kavssery P and Lips, Alex},
  journal={Langmuir},
  volume={22},
  number={18},
  pages={7528--7542},
  year={2006},
  publisher={ACS Publications}
}

@article{garrett1989reexamination,
  title={A reexamination of the measurement of dynamic surface tensions using the maximum bubble pressure method},
  author={Garrett, Peter R and Ward, David R},
  journal={Journal of Colloid and Interface Science},
  volume={132},
  number={2},
  pages={475--490},
  year={1989},
  publisher={Elsevier}
}

@article{mishchuk2001hydrodynamic,
  title={Hydrodynamic processes in dynamic bubble pressure experiments: Part 5. The adsorption at the surface of a growing bubble},
  author={Mishchuk, NA and Dukhin, SS and Fainerman, VB and Kovalchuk, VI and Miller, R},
  journal={Colloids and Surfaces A: Physicochemical and Engineering Aspects},
  volume={192},
  number={1-3},
  pages={157--175},
  year={2001},
  publisher={Elsevier}
}

@article{berry2015measurement,
  title={Measurement of surface and interfacial tension using pendant drop tensiometry},
  author={Berry, Joseph D and Neeson, Michael J and Dagastine, Raymond R and Chan, Derek YC and Tabor, Rico F},
  journal={Journal of colloid and interface science},
  volume={454},
  pages={226--237},
  year={2015},
  publisher={Elsevier}
}

@article{reichert2015importance,
  title={The importance of experimental design on measurement of dynamic interfacial tension and interfacial rheology in diffusion-limited surfactant systems},
  author={Reichert, Matthew D and Alvarez, Nicolas J and Brooks, Carlton F and Grillet, Anne M and Mondy, Lisa A and Anna, Shelley L and Walker, Lynn M},
  journal={Colloids and Surfaces A: Physicochemical and Engineering Aspects},
  volume={467},
  pages={135--142},
  year={2015},
  publisher={Elsevier}
}

@article{WARD.1946,
 author = {Ward, A. F. H. and Tordai, L.},
 year = {1946},
 title = {Time--Dependence of Boundary Tensions of Solutions I. The Role of Diffusion in Time--Effects},
 pages = {453--461},
 volume = {14},
 number = {7},
 issn = {0021-9606},
 journal = {The Journal of Chemical Physics},
 doi = {10.1063/1.1724167}
}

@article{WARD.1944,
 author = {Ward, A. F. H. and Tordai, L.},
 year = {1944},
 title = {Existence of Time-Dependence for Interfacial Tension of Solutions},
 url = {https://www.nature.com/articles/154146b0},
 pages = {146--147},
 volume = {154},
 number = {3900},
 issn = {1476-4687},
 journal = {Nature},
 doi = {10.1038/154146b0}
}

@article{Kovalchuk.2023,
 author = {Kovalchuk, Nina M. and Simmons, Mark J. H.},
 year = {2023},
 title = {Review of the role of surfactant dynamics in drop microfluidics},
 url = {https://www.sciencedirect.com/science/article/pii/S0001868623000118},
 pages = {102844},
 volume = {312},
 issn = {0001-8686},
 journal = {Advances in Colloid and Interface Science},
 doi = {10.1016/j.cis.2023.102844}
}

@article{Deng.2022,
 author = {Deng, Boxin and Schro{\"e}n, Karin and de Ruiter, Jolet},
 year = {2022},
 title = {Dynamics of bubble formation in spontaneous microfluidic devices: Controlling dynamic adsorption via liquid phase properties},
 url = {https://www.sciencedirect.com/science/article/pii/S0021979722006841},
 pages = {218--227},
 volume = {622},
 issn = {0021-9797},
 journal = {Journal of Colloid and Interface Science},
 doi = {10.1016/j.jcis.2022.04.115}
}

@article{Liang.2022,
 author = {Liang, Xiao and Zhang, Jiyizhe and Li, Min and Wang, Kai and Luo, Guangsheng},
 year = {2022},
 title = {Dynamic interfacial tension and adsorption kinetics of nonionic surfactants during microfluidic droplet formation process},
 urldate = {13/12/2022},
 pages = {136658},
 volume = {445},
 issn = {13858947},
 journal = {Chemical Engineering Journal},
 doi = {10.1016/j.cej.2022.136658}
}

@article{Alvarez.2010micro,
 author = {Alvarez, Nicolas J. and Walker, Lynn M. and Anna, Shelley L.},
 year = {2010},
 title = {A microtensiometer to probe the effect of radius of curvature on surfactant transport to a spherical interface},
 pages = {13310--13319},
 volume = {26},
 number = {16},
 journal = {Langmuir : the ACS journal of surfaces and colloids},
 doi = {10.1021/la101870m}
}

@article{Brosseau.2014,
 author = {Brosseau, Quentin and Vrignon, J{\'e}r{\'e}my and Baret, Jean-Christophe},
 year = {2014},
 title = {Microfluidic Dynamic Interfacial Tensiometry ($\mu$DIT)},
 url = {https://pubs.rsc.org/en/content/articlehtml/2014/sm/c3sm52543k},
 pages = {3066--3076},
 volume = {10},
 number = {17},
 journal = {Soft Matter},
 doi = {10.1039/C3SM52543K}
}

@article{nele2022analytically,
  title={An analytically solvable numerical approximation to Ward-Tordai equation applied to Langmuir adsorption},
  author={Nele, M{\'a}rcio},
  journal={Colloids and Surfaces A: Physicochemical and Engineering Aspects},
  volume={648},
  pages={129186},
  year={2022},
  publisher={Elsevier}
}

@article{lin1990diffusion,
  title={Diffusion-controlled surfactant adsorption studied by pendant drop digitization},
  author={Lin, Shi-Yow and McKeigue, Kevin and Maldarelli, Charles},
  journal={AIChE Journal},
  volume={36},
  number={12},
  pages={1785--1795},
  year={1990},
  publisher={Wiley Online Library}
}

@article{kovalchuk2023surfactant,
  title={Surfactant Adsorption Layers: Experiments and Modeling},
  author={Kovalchuk, VI and Aksenenko, EV and Schneck, E and Miller, R},
  journal={Langmuir},
  volume={39},
  number={10},
  pages={3537--3545},
  year={2023},
  publisher={ACS Publications}
}

\end{document}


\title{\textbf{Supporting Information for} \\ Geometry-Controlled Dynamic Tensiometry Resolves Intrinsic Surfactant Adsorption Kinetics}
\author{Camille Brigodiot\(^{a}\)\footnote{Corresponding author. \\E-mail adress: c.brigodiot@uva.nl}, Boxin Deng\(^b\), Christine Dalmazzone\(^c\),\\ Karin Schroën\(^b\), Annie Colin\(^d\)
\and a: Van der Waals - Zeeman Institute, Institute of Physics, University of Amsterdam,\\ Science Park 904, 1098 XH Amsterdam, The Netherlands \\
b: Wageningen University and Research, Laboratory of Food Process Engineering,\\ Bornse Weilanden 9, 6708 WG Wageningen, the Netherlands \\
c: 	IFP Energies nouvelles, 1 et 4 avenue de Bois-Préau, 92852 Rueil-Malmaison, France \\
d: MIE, CBI, ESPCI Paris, Université PSL, CNRS 75005 Paris, France }
\date{}
\maketitle

\section{EDGE geometry}

\begin{figure}[ht]
    \centering
    \includegraphics[width=\linewidth]{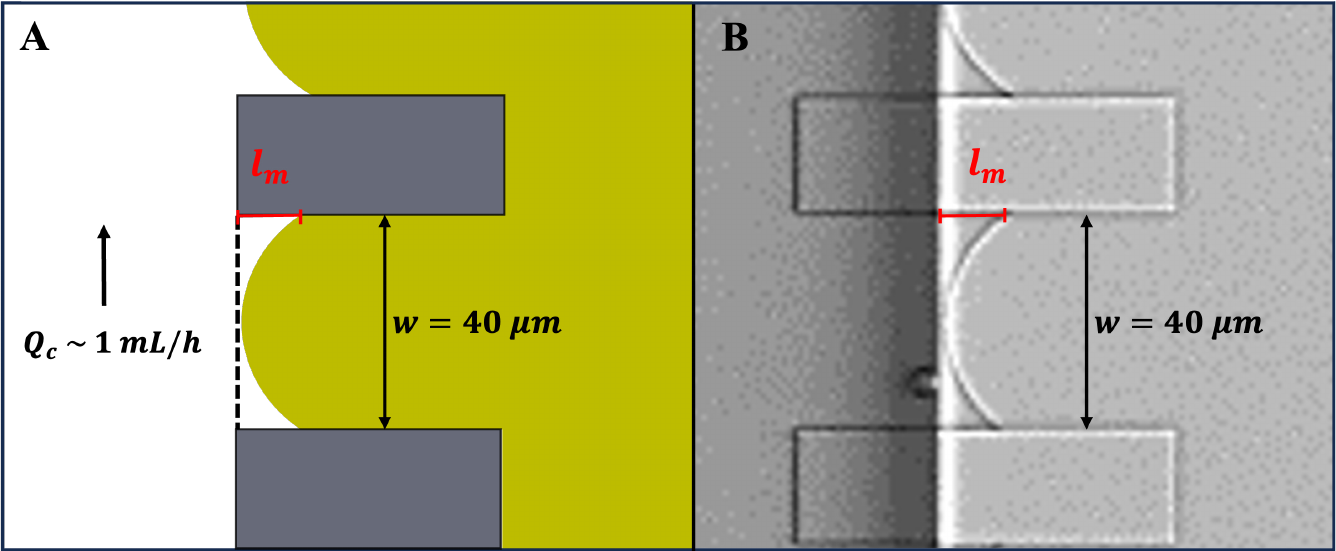}
    \caption{Scheme (A) and picture (B) of the oil/water interface at one pore. The dispersed phase is oil and the continuous phase is water, with an average flow rate of 1 \si{\milli \liter \per \hour}. In (A), the dotted line indicates the edge of the pore. The scale is given by the width of the pore w= 40 \si{\micro \meter}.}
    \label{fig:lm}
\end{figure}

Figure~\ref{fig:lm} shows that the meniscus is curved both across the shallow pore height and along the pore width $W$. The feeding area is defined as the planar rectangular projection
\begin{equation}
A_f=Wh,
\end{equation}
whereas $A_i$ is the actual curved oil--water interfacial area.

Across the pore height, the meniscus is approximated by a circular arc of radius $R$. If $\theta$ is the contact angle, the half-angle of the arc is:
\begin{equation}
\alpha=\frac{\pi}{2}-\theta,
\end{equation}
and the pore height is related to the radius through
\begin{equation}
h=2R\sin\alpha=2R\cos\theta.
\end{equation}
The lateral recession of the meniscus along the pore width is described
by the parabolic profile
\begin{equation}
X(y)=\ell_m\left(\frac{2y}{W}\right)^2,\qquad
-\frac{W}{2}\leq y\leq\frac{W}{2}.
\end{equation}

The complete meniscus surface can therefore be parametrized as:
\begin{equation}
\mathbf r(y,\varphi)= \left(X(y)+R[\cos\varphi-\cos\alpha],\,
y,\, R\sin\varphi
\right),
\end{equation}
with
\begin{equation}
-\frac{W}{2}\leq y\leq\frac{W}{2},\qquad
-\alpha\leq\varphi\leq\alpha.
\end{equation}
The corresponding surface element is
\begin{equation}
\mathrm dA=R\sqrt{1+\left(\frac{\mathrm dX}{\mathrm dy}\right)^2 \cos^2\varphi}\,\mathrm dy\,\mathrm d\varphi,
\end{equation}
where
\begin{equation}
\frac{\mathrm dX}{\mathrm dy}=\frac{8\ell_m y}{W^2}.
\end{equation}
The actual interfacial area is consequently
\begin{equation}
A_i=\int_{-W/2}^{W/2}\int_{-\alpha}^{\alpha}R\sqrt{ 1+\left(\frac{8\ell_m y}{W^2}\right)^2\cos^2\varphi} \,\mathrm d\varphi\,\mathrm dy.
\end{equation}
Using $A_f=Wh=2RW\sin\alpha$ and introducing $t=2y/W$, the area ratio becomes
\begin{equation}
\frac{A_i}{A_f}= \frac{1}{\sin\alpha} \int_0^1 \int_0^\alpha \sqrt{ 1+ \left(\frac{4\ell_m}{W}\right)^2 t^2\cos^2\varphi} \,\mathrm d\varphi\,\mathrm dt.
\end{equation}
For
\begin{equation}
W=40\si{\micro \meter},
\qquad
\ell_m=12\si{\micro \meter},
\qquad
\theta=31.8^\circ,
\end{equation}
one has
\begin{equation}
\alpha=1.016~\mathrm{rad},
\qquad
\frac{4\ell_m}{W}=1.2.
\end{equation}
Numerical evaluation of the double integral gives
\begin{equation}
\boxed{
\frac{A_i}{A_f}=1.376\simeq1.38.
}
\end{equation}
This geometrical factor accounts simultaneously for the curvature
across the pore height and for the lateral recession of the meniscus
along the pore width.

\section{Characteristic scales}
In EDGE, surfactant transport towards the interface is described by the effective diffusion length $L_D$ calculated directly from the device geometry and flow conditions.At transverse position $y$, the transport distance combines the convective depletion layer and the recessed meniscus depth,
\begin{equation}
d(y,U)=\left[\frac{\pi D(y+w/2)}{U}\right]^{1/2}+\ell_m\left(\frac{2y}{w}\right)^2.
\end{equation}
Averaging the corresponding local conductance over the meniscus width gives

\begin{equation}
L_D=\frac{w}{\displaystyle\int_{-w/2}^{w/2}\frac{dy}{d(y,U)}}.
\label{eq:LD_EDGE}
\end{equation}

Thus, $L_D$ is not an adjustable diffusion distance but a geometry- and flow-dependent transport length that remains in the micrometre range under the present EDGE conditions.
Denoting by $A_f$ the area of the renewed liquid film feeding the meniscus and by $A_i$ the interfacial area, the average diffusive flux per unit interfacial area is written as:

\begin{equation}
J_D^{\mathrm{EDGE}}
=
\frac{A_f}{A_i}
\frac{D}{L_D}
\left(c_b-c_s\right),
\label{eq:edge_diffusive_flux}
\end{equation}
where $c_s$ is the subsurface concentration at the meniscus.
Unlike diffusion towards a planar semi-infinite interface, the EDGE flux
does not decay as $t^{-1/2}$ within this reduced transport description.
The transport distance is fixed by the geometry and flow conditions through
$L_D$.

For an initially clean interface behaving as a perfect sink,
$c_s\simeq 0$, the amount supplied to the interface therefore increases
linearly with time:
\begin{equation}
\Gamma_D^{\mathrm{EDGE}}(t)
\simeq
\frac{A_f}{A_i}
\frac{D c_b}{L_D}t.
\label{eq:edge_short_time_supply}
\end{equation}

The time required to supply a prescribed surface excess $\Gamma^*$ is
therefore
\begin{equation}
\tau_{D,\mathrm{sup}}^{\mathrm{EDGE}}
\left(\Gamma^*,c_b\right)
=
\frac{A_i}{A_f}
\frac{\Gamma^* L_D}{D c_b}.
\label{eq:edge_supply_time}
\end{equation}

For a prescribed surface excess, the EDGE supply time consequently scales as
\begin{equation}
\tau_{D,\mathrm{sup}}^{\mathrm{EDGE}}
\propto c_b^{-1},
\end{equation}
rather than as $c_b^{-2}$ for planar semi-infinite diffusion. This change in
concentration scaling results directly from the fixed micrometre-scale
transport distance imposed by EDGE.

If the target surface excess is taken as the Langmuir equilibrium value ,
\begin{equation}
\Gamma_{\mathrm{eq}}
=
\Gamma_\infty
\frac{Kc_b}{1+Kc_b},
\end{equation}
Equation~\eqref{eq:edge_supply_time} becomes
\begin{equation}
\tau_{D,\mathrm{sup}}^{\mathrm{EDGE}}
=
\frac{A_i}{A_f}
\frac{\Gamma_\infty K L_D}
{D(1+Kc_b)}.
\label{eq:edge_supply_time_equilibrium}
\end{equation}

At low concentration, $Kc_b\ll 1$, the amount adsorbed at equilibrium is
itself proportional to $c_b$, and the equilibrium supply time therefore
approaches a concentration-independent value. At high concentration,
$Kc_b\gg 1$, the interface approaches saturation and
$\tau_{D,\mathrm{sup}}^{\mathrm{EDGE}}\propto c_b^{-1}$.

\paragraph{Nonionic surfactant.}

For a nonionic surfactant, the diffusion-controlled limit assumes that the
interface remains in local equilibrium with the subsurface concentration:
\begin{equation}
\phi(t)
=
\phi_{\mathrm{eq}}\!\left(c_s(t)\right)
=
\frac{Kc_s(t)}{1+Kc_s(t)}.
\label{eq:edge_langmuir_local_equilibrium}
\end{equation}

The interfacial material balance is then
\begin{equation}
\Gamma_\infty
\frac{d\phi}{dt}
=
\frac{A_f}{A_i}
\frac{D}{L_D}
\left(c_b-c_s\right).
\label{eq:edge_diffusion_only_balance}
\end{equation}

Using
\begin{equation}
\frac{d\phi_{\mathrm{eq}}}{dc_s}
=
\frac{K}{(1+Kc_s)^2},
\end{equation}
the equation solved for $c_s(t)$ is
\begin{equation}
\frac{dc_s}{dt}
=
\frac{A_f}{A_i}
\frac{D}{\Gamma_\infty L_D}
\frac{(1+Kc_s)^2}{K}
\left(c_b-c_s\right).
\label{eq:edge_diffusion_only_cs}
\end{equation}

Because the Langmuir isotherm is nonlinear, the local diffusive relaxation
time changes during the tension curve. Around a given subsurface
concentration $c_s^*$, it is
\begin{equation}
\tau_D^{\mathrm{EDGE}}(c_s^*)
=
\frac{A_i}{A_f}
\frac{\Gamma_\infty L_D}{D}
\left.
\frac{d\phi_{\mathrm{eq}}}{dc_s}
\right|_{c_s=c_s^*}.
\label{eq:edge_local_diffusion_time}
\end{equation}

At the beginning of the curve, $c_s\simeq 0$, and
\begin{equation}
\boxed{
\tau_{D,0}^{\mathrm{N}}
=
\frac{A_i}{A_f}
\frac{\Gamma_\infty K L_D}{D}.
}
\label{eq:edge_initial_diffusion_time}
\end{equation}

Although this initial local time is independent of $c_b$, the initial
adsorption rate is not. Indeed,
\begin{equation}
\left.
\frac{d\phi}{dt}
\right|_{t=0}
=
\frac{A_f}{A_i}
\frac{D c_b}{\Gamma_\infty L_D},
\label{eq:edge_initial_diffusion_slope}
\end{equation}
and therefore increases linearly with bulk concentration.

Near equilibrium, $c_s\simeq c_b$, the diffusive relaxation time becomes
\begin{equation}
\boxed{
\tau_{D,\mathrm{eq}}^{\mathrm{N}}(c_b)
=
\frac{A_i}{A_f}
\frac{\Gamma_\infty K L_D}
{D(1+Kc_b)^2}.
}
\label{eq:edge_final_diffusion_time}
\end{equation}

The difference between
$\tau_{D,0}^{\mathrm{N}}$ and
$\tau_{D,\mathrm{eq}}^{\mathrm{N}}$
arises from the differential interfacial capacity
\begin{equation}
\Gamma_\infty
\frac{d\phi_{\mathrm{eq}}}{dc_s}
=
\frac{\Gamma_\infty K}{(1+Kc_s)^2},
\end{equation}
which decreases as the interface approaches saturation.

In the adsorption-controlled limit, transport maintains $c_s\simeq c_b$ and
the coverage evolves as
\begin{equation}
\frac{d\phi}{dt}
=
k_a c_b(1-\phi)-k_d\phi.
\label{eq:edge_nonionic_adsorption}
\end{equation}

The corresponding molecular adsorption time is
\begin{equation}
\boxed{
\tau_A^{\mathrm{N}}(c_b)
=
\frac{1}{k_a c_b+k_d}.
}
\label{eq:edge_nonionic_adsorption_time}
\end{equation}

Using $K=k_a/k_d$, a geometry-dependent EDGE Damk\"ohler number may be
defined as
\begin{equation}
\boxed{
\mathrm{Da}_{\mathrm{EDGE}}
=
\frac{A_i}{A_f}
\frac{\Gamma_\infty k_a L_D}{D}.
}
\label{eq:edge_damkohler}
\end{equation}

At the beginning of the curve, the ratio between the local diffusion and
adsorption times is
\begin{equation}
\frac{\tau_{D,0}^{\mathrm{N}}}
{\tau_A^{\mathrm{N}}}
=
\mathrm{Da}_{\mathrm{EDGE}}
\left(1+Kc_b\right),
\label{eq:edge_initial_time_ratio}
\end{equation}
whereas near equilibrium it is
\begin{equation}
\frac{\tau_{D,\mathrm{eq}}^{\mathrm{N}}}
{\tau_A^{\mathrm{N}}}
=
\frac{\mathrm{Da}_{\mathrm{EDGE}}}
{1+Kc_b}.
\label{eq:edge_final_time_ratio}
\end{equation}

At low concentration, $Kc_b\ll1$, the two ratios approach the same
geometry-dependent limit:
\begin{equation}
\frac{\tau_D}{\tau_A}
\simeq
\mathrm{Da}_{\mathrm{EDGE}}.
\end{equation}

Consequently, dilution does not force EDGE into a diffusion-controlled regime. Instead, the low-concentration mechanism is selected primarily by the fixed geometry-dependent number $\mathrm{Da}_{\mathrm{EDGE}}$: adsorption is slower when $\mathrm{Da}_{\mathrm{EDGE}}<1$, whereas transport is slower when $\mathrm{Da}_{\mathrm{EDGE}}>1$.

At higher concentrations, the mechanism may evolve during a single tension curve. Transport can limit the initial response because the molecular adsorption rate increases with $c_b$, whereas the final relaxation can become comparatively more sensitive to adsorption because the differential surface capacity decreases as saturation is approached.

\paragraph{Ionic surfactant.}

For an ionic surfactant, the EDGE transport equation remains unchanged, but
the molecular adsorption rate is reduced by the electrostatic potential
created by the charged interface. Introducing the dimensionless
electrostatic barrier
\begin{equation}
B(\phi)
=
\frac{ze\psi_0(\phi)}{k_{\mathrm B}T},
\end{equation}
with the sign convention that $B>0$ corresponds to electrostatic repulsion,
the adsorption rate is written as
\begin{equation}
R_{\mathrm{ads}}^{\mathrm{I}}(c_s,\phi)
=
k_a c_s(1-\phi)\exp[-B(\phi)]
-k_d\phi.
\label{eq:edge_ionic_adsorption_rate}
\end{equation}

In the adsorption-controlled limit, $c_s\simeq c_b$, and
\begin{equation}
\frac{d\phi}{dt}
=
R_{\mathrm{ads}}^{\mathrm{I}}(c_b,\phi).
\end{equation}

Because the electrostatic barrier depends on coverage, the ionic dynamics
are nonlinear. The molecular relaxation time around the equilibrium
coverage $\phi_{\mathrm{eq}}$ is
\begin{equation}
\tau_A^{\mathrm{I}}
=
-
\left[
\left.
\frac{\partial R_{\mathrm{ads}}^{\mathrm{I}}}
{\partial\phi}
\right|_{c_s=c_b,\phi=\phi_{\mathrm{eq}}}
\right]^{-1}.
\label{eq:edge_ionic_adsorption_time_general}
\end{equation}

For Equation~\eqref{eq:edge_ionic_adsorption_rate}, this gives
\begin{equation}
\boxed{
\left(\tau_A^{\mathrm{I}}\right)^{-1}
=
k_d
+
k_a c_b
\exp(-B_{\mathrm{eq}})
\left[
1+
(1-\phi_{\mathrm{eq}})
B'_{\mathrm{eq}}
\right],
}
\label{eq:edge_ionic_adsorption_time}
\end{equation}
where
\begin{equation}
B_{\mathrm{eq}}=B(\phi_{\mathrm{eq}}),
\qquad
B'_{\mathrm{eq}}
=
\left.
\frac{dB}{d\phi}
\right|_{\phi=\phi_{\mathrm{eq}}}.
\end{equation}

In the diffusion-controlled limit, the interface is assumed to remain in
local electrostatic equilibrium with $c_s$. The ionic equilibrium isotherm
is defined implicitly by
\begin{equation}
R_{\mathrm{ads}}^{\mathrm{I}}
\left(c_s,\phi_{\mathrm{eq}}^{\mathrm{I}}(c_s)\right)
=0.
\label{eq:edge_ionic_equilibrium}
\end{equation}

The corresponding diffusion-only equation is
\begin{equation}
\Gamma_\infty
\frac{d\phi_{\mathrm{eq}}^{\mathrm{I}}}{dc_s}
\frac{dc_s}{dt}
=
\frac{A_f}{A_i}
\frac{D}{L_D}
(c_b-c_s).
\label{eq:edge_ionic_diffusion_only}
\end{equation}

The local ionic diffusion time around a concentration $c_s^*$ is therefore
\begin{equation}
\boxed{
\tau_D^{\mathrm{I}}(c_s^*)
=
\frac{A_i}{A_f}
\frac{\Gamma_\infty L_D}{D}
\left.
\frac{d\phi_{\mathrm{eq}}^{\mathrm{I}}}{dc_s}
\right|_{c_s=c_s^*}.
}
\label{eq:edge_ionic_diffusion_time_general}
\end{equation}

Implicit differentiation of
Equation~\eqref{eq:edge_ionic_equilibrium} gives
\begin{equation}
\frac{d\phi_{\mathrm{eq}}^{\mathrm{I}}}{dc_s}
=
-
\frac{
\partial R_{\mathrm{ads}}^{\mathrm{I}}/\partial c_s
}{
\partial R_{\mathrm{ads}}^{\mathrm{I}}/\partial\phi
}.
\label{eq:edge_ionic_isotherm_derivative_general}
\end{equation}

For the kinetic law in
Equation~\eqref{eq:edge_ionic_adsorption_rate},
\begin{equation}
\frac{d\phi_{\mathrm{eq}}^{\mathrm{I}}}{dc_s}
=
\frac{
k_a(1-\phi_{\mathrm{eq}})
\exp(-B_{\mathrm{eq}})
}{
k_d
+
k_a c_s\exp(-B_{\mathrm{eq}})
\left[
1+
(1-\phi_{\mathrm{eq}})B'_{\mathrm{eq}}
\right]
}.
\label{eq:edge_ionic_isotherm_derivative}
\end{equation}

At the beginning of the curve, $\phi\simeq0$. If the clean interface is
uncharged, $B(0)=0$, and the dilute-limit equilibrium slope reduces to
\begin{equation}
\left.
\frac{d\phi_{\mathrm{eq}}^{\mathrm{I}}}{dc_s}
\right|_{c_s=0}
=
\frac{k_a}{k_d}
=
K.
\end{equation}

The initial ionic diffusion time is consequently identical to the nonionic
one:
\begin{equation}
\boxed{
\tau_{D,0}^{\mathrm{I}}
=
\frac{A_i}{A_f}
\frac{\Gamma_\infty K L_D}{D}.
}
\label{eq:edge_ionic_initial_diffusion_time}
\end{equation}

Electrostatic effects emerge as the interface becomes charged. Near
equilibrium,
\begin{equation}
\boxed{
\tau_{D,\mathrm{eq}}^{\mathrm{I}}
=
\frac{A_i}{A_f}
\frac{\Gamma_\infty L_D}{D}
\frac{
k_a(1-\phi_{\mathrm{eq}})
\exp(-B_{\mathrm{eq}})
}{
k_d
+
k_a c_b\exp(-B_{\mathrm{eq}})
\left[
1+
(1-\phi_{\mathrm{eq}})B'_{\mathrm{eq}}
\right]
}.
}
\label{eq:edge_ionic_final_diffusion_time}
\end{equation}

Combining Equations~\eqref{eq:edge_ionic_adsorption_time} and
\eqref{eq:edge_ionic_final_diffusion_time} yields the particularly simple
ratio
\begin{equation}
\boxed{
\frac{\tau_{D,\mathrm{eq}}^{\mathrm{I}}}
{\tau_A^{\mathrm{I}}}
=
\mathrm{Da}_{\mathrm{EDGE}}
(1-\phi_{\mathrm{eq}})
\exp(-B_{\mathrm{eq}}).
}
\label{eq:edge_ionic_time_ratio}
\end{equation}

In the absence of electrostatic interactions,
$B_{\mathrm{eq}}=0$, and
$1-\phi_{\mathrm{eq}}=(1+Kc_b)^{-1}$, so
Equation~\eqref{eq:edge_ionic_time_ratio} reduces to the nonionic result in
Equation~\eqref{eq:edge_final_time_ratio}.

At weak ionic strength, the electrostatic barrier increases as the interface is charged. The factor $\exp(-B_{\mathrm{eq}})$ reduces the sensitivity of the adsorption rate to the subsurface concentration and shifts the dynamics towards adsorption control. Adding electrolyte screens the surface charge, decreases $B_{\mathrm{eq}}$, and progressively restores the nonionic EDGE limit.

Thus, EDGE does not eliminate transport--adsorption coupling. Instead, its fixed and independently calculable transport length makes that coupling quantifiable. At low concentration, the balance is governed primarily by $\mathrm{Da}_{\mathrm{EDGE}}$, while at higher coverage the nonlinear interfacial capacity and, for ionic surfactants, the electrostatic adsorption barrier determine how the controlling mechanism evolves along the tension curve.

\section{Analysis of the experimental data for a non-ionic surfactant limited solely by adsorption} 

In this section, we aim to analyse the hypothesis of a naked interface at t=0. The analysis is done in the case of $\mathrm{C_{10}E_8}$ that is limited solely  by adsorption in the EDGE device. Equation 7 from the main text, coupled with Equation 10 are used. 
We obtain  for $\beta=0$ the following expression:

\begin{equation}
    \gamma(t) - \gamma_0= \frac{1}{a^2} \Big[ k_BT\Big(\phi_0(t) 
ln(\phi_0(t))+(1-\phi_0(t))ln(1-\phi_0(t)) - \phi_0(t)ln(c_ba^3N_A) \Big) 
-\alpha \phi_0 \Big]\\
\label{eq:gamma_t}
\end{equation}

With : 
\begin{equation}
\phi_0(t)=(\phi_{eq}-\phi_{0}^{t_{drop}}(0))*(1-exp^{(-t/\tau)}+ \phi_{0}^{t_{drop}}(0)
 \end{equation}

To ensure clearer notation and for pedagogical reasons, we have chosen to denote: $\phi_0(0)=\phi_{0}^{t_{drop}}(0)$, where $t_drop$ is the droplet formation time as defined in the previous section.  
Indeed $\phi_{0}(0)$ in our experiment is a constant  for a given serie of formed droplet ie for a given pressure and a given surfactant concentration but it may change with each applied pressure and, therefore, with each experimental time point or $t_{drop}$ . In fact, even if it is the value  at $t=0$, it is actually a function that depends on $t$ as this value depends upon each pressure drop.
Fitting this equation to the  experimental data points thus amounts to determining these X values of $\phi_{0}^{t_{drop}}(0)$ together with the value of $\tau$. A priori, the system therefore contains more unknowns than equations; it is underdetermined. However, an important constraint remains: the values of $\phi_{0}^{t_{drop}} (0)$ must be positive, and they must increase with experimental time. 

We must therefore calculate the initial coverage rates for each concentration. 
To do this, we will adopt the following approach: we will fix \(\tau\) and calculate the initial coverage rates $\phi_{0}^{t_{drop}}(0)$,  and then assess whether they are physically reasonable. Let us first focus on the most diluted concentration (c=0.01 CMC). 

\begin{figure}[ht]
    \centering
    \includegraphics[width=1\linewidth]{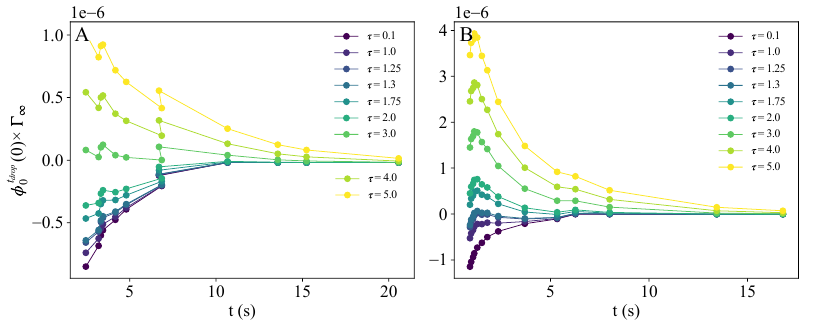}
    \caption{Assessment of the naked interface hypothesis. (A) Calculus of $\mathrm{\phi_0^{t_{drop}}\times \Gamma_\infty}$ for c=0.01 CMC at different fixed $\tau$ for $\mathrm{C_{10}E_8}$. (B) Calculus of $\mathrm{\phi_0^{t_{drop}}\times \Gamma_\infty}$ for c=0.025 CMC at different fixed $\tau$ for $\mathrm{C_{10}E_8}$.}
    \label{fig:phi0}
\end{figure}

As shown in \autoref{fig:phi0}, we calculate the initial coverage rates for different droplet formation times. When \(\tau\) is greater than 5 seconds or less than 3, we obtain results that are not noisy but physically unrealistic: either the values are negative when \(\tau\) is less than 2 s, or the values decrease with the droplet formation time when \(\tau\) is greater than 4 s. The model therefore only makes sense if \(\tau\) is between 2 and 4. 
Within this framework, the initial coverage rates are low and can be neglected. It is thus possible to set them to zero and to associate the droplet formation time with the interface lifetime, from the moment when it is bare to the instant when it is covered by surfactant.
This analysis can be performed for all the concentrations. It shows that within the range of \(\tau\) values for which the model makes sense, the initial rates of adsorbed surfactant were necessarily equal to zero. We will keep this hypothesis for the next steps.

\section{Electrostatic contribution for ionic surfactants}

For the ionic surfactant, the interfacial thermodynamics were described by combining the nonequilibrium Diamant--Andelman \cite{diamant1996kinetics, diamant2001} free energy with a Gouy--Chapman description of \cite{levine1963discrete, bonfillon1994dynamic, Prosser.2001.review} the diffuse ionic layer. The electrostatic contribution enters separately in the adsorption equilibrium and in the interfacial tension.

\subsection{Surface charge and Gouy--Chapman potential}

The surface concentration of adsorbed surfactant is written as

\begin{equation}
\Gamma=\Gamma_{\infty}\phi,
\end{equation}

where $\Gamma_{\infty}$ is the maximum surface excess and $\phi$ is the fractional surface coverage. The magnitude of the nominal surface charge density is then

\begin{equation}
\sigma = F\Gamma_{\infty}\phi,
\end{equation}

where $F$ is Faraday's constant.

Within the Gouy--Chapman description for a monovalent electrolyte, we introduce

\begin{equation}
A(c_s) = \sqrt{8\varepsilon_r\varepsilon_0 RTc_s},
\end{equation}

where $c_s$ is the surfactant concentration in the liquid immediately adjacent to the interface, $\varepsilon_r$ is the relative permittivity of water, and $\varepsilon_0$ is the vacuum permittivity.

The dimensionless Gouy--Chapman variable is

\begin{equation}
y = \operatorname{asinh} \left[ \frac{\sigma}{A(c_s)} \right],
\end{equation}

and the magnitude of the surface potential is therefore

\begin{equation}
|\psi_0| = \frac{2RT}{F} \operatorname{asinh} \left[ \frac{F\Gamma_{\infty}\phi} {\sqrt{8\varepsilon_r\varepsilon_0 RTc_s}}\right].
\label{eq:SI_psi0}
\end{equation}

Equivalently, the electrostatic energy associated with bringing one mole of charged surfactant to the interface can be written in dimensionless form as

\begin{equation}
u_{\mathrm{el}} = \frac{F|\psi_0|}{RT}.
\end{equation}

\subsection{Electrostatic correction to the adsorption isotherm}

At equilibrium, $c_s=c_b$ and $\phi=\phi_{\mathrm{eq}}$. The adsorption isotherm is written as:

\begin{equation}
\frac{\phi_{\mathrm{eq}}} {1-\phi_{\mathrm{eq}}}
= Kc_b \exp \left[ \beta^\star\phi_{\mathrm{eq}}-
\chi_{\mathrm{el}} \frac{F|\psi_0|}{RT} \right].
\label{eq:SI_ionic_isotherm}
\end{equation}

Here, $K$ is the adsorption equilibrium constant, $\beta^\star$ is the dimensionless lateral-interaction parameter, and $\chi_{\mathrm{el}}$ is an effective electrostatic weighting factor. The factor $\chi_{\mathrm{el}}$ accounts for the fact that the effective electrostatic penalty experienced by the adsorbing surfactant may be smaller than that obtained from the nominal surface charge, for example because of counterion association.

In the analysis reported here, the lateral interaction parameter is fixed to $\beta^\star=0$, so that Eq.~\eqref{eq:SI_ionic_isotherm} reduces to the following:

\begin{equation}
\frac{\phi_{\mathrm{eq}}} {1-\phi_{\mathrm{eq}}}
= Kc_b \exp \left[ -
\chi_{\mathrm{el}} \frac{F|\psi_0|}{RT} \right].
\label{eq:SI_ionic_isotherm_beta0}
\end{equation}

Thus, the departure from a Langmuir isotherm arises here solely from the electrostatic cost of charging the interface.

\subsection{Chemical contribution to the interfacial tension}

For a given subsurface concentration $c_s$ and surface coverage $\phi$, the nonequilibrium chemical contribution to the interfacial tension is

\begin{equation}
\begin{split}
\gamma_{\mathrm{chem}}(c_s,\phi)-\gamma_0
= RT\Gamma_{\infty} \bigg[
& \phi\ln\phi +(1-\phi)\ln(1-\phi)
\\
&-\phi\ln(N_Aa^3c_s) -\alpha^\star\phi -\frac{\beta^\star}{2}\phi^2 \bigg],
\end{split}
\label{eq:SI_gamma_chem}
\end{equation}

where $a$ is the molecular length scale and $\alpha^\star=\alpha/(k_{\mathrm B}T)$ is the dimensionless adsorption affinity. The molecular length is related to the limiting surface excess through:

\begin{equation}
a= \left( \frac{1}{N_A\Gamma_{\infty}} \right)^{1/2},
\end{equation}

and the adsorption affinity is obtained from

\begin{equation}
\alpha^\star = \ln \left( \frac{K}{N_Aa^3} \right).
\end{equation}

For the present SDS analysis, $\beta^\star=0$ is imposed in Eq.~\eqref{eq:SI_gamma_chem}.

\subsection{Diffuse-layer contribution to the interfacial tension}
The Gouy--Chapman free energy per unit area associated with charging the diffuse layer is calculated as: 

\begin{equation}
g_{\mathrm{GC}} = \frac{2RT}{F} A(c_s) \left[ y\sinh y-\cosh y+1 \right],
\label{eq:SI_GC_energy}
\end{equation}

with

\begin{equation}
y= \operatorname{asinh} \left[ \frac{F\Gamma_{\infty}\phi} {\sqrt{8\varepsilon_r\varepsilon_0RTc_s}} \right].
\end{equation}

The electrostatic contribution to the measured interfacial tension is then written as:

\begin{equation}
\gamma_{\mathrm{el}}(c_s,\phi)
=
\chi_{\gamma}g_{\mathrm{GC}}(c_s,\phi),
\label{eq:SI_gamma_el}
\end{equation}

and the total interfacial tension is

\begin{equation}
\boxed{
\gamma(c_s,\phi)
=
\gamma_{\mathrm{chem}}(c_s,\phi)
+
\gamma_{\mathrm{el}}(c_s,\phi)
}.
\label{eq:SI_gamma_total}
\end{equation}

The parameters $\chi_{\mathrm{el}}$ and $\chi_\gamma$ have distinct
roles. The former weights the electrostatic penalty entering the
adsorption equilibrium and adsorption kinetics, whereas the latter
weights the contribution of diffuse-layer charging to the measured
interfacial tension. They are therefore not required to be identical.

With the sign convention used here, $g_{\mathrm{GC}}\geq0$. A positive
$\chi_\gamma$ consequently corresponds to a positive electrostatic
contribution to the interfacial free energy: charging the interface
partially opposes the decrease in interfacial tension produced by
surfactant adsorption.

\subsection{Determination of the equilibrium parameters}

For each bulk concentration, Eq.~\eqref{eq:SI_ionic_isotherm_beta0} is
solved self-consistently because $\psi_0$ depends on the unknown
equilibrium coverage $\phi_{\mathrm{eq}}$. The resulting coverage is then
inserted into Eqs.~\eqref{eq:SI_gamma_chem}--\eqref{eq:SI_gamma_total}
to calculate the equilibrium interfacial tension.

In the present analysis, $\chi_{\mathrm{el}}$ is fixed independently,
while $\Gamma_{\infty}$, $K$, and $\chi_\gamma$ are determined from the
equilibrium interfacial-tension data. All these parameters are subsequently
kept fixed when analyzing the dynamic EDGE measurements.

\section{Nonionic surfactant: dynamic regimes}

\subsection{Adsorption regime}
\begin{figure}[ht]
    \centering
    \includegraphics[width=0.7\linewidth]{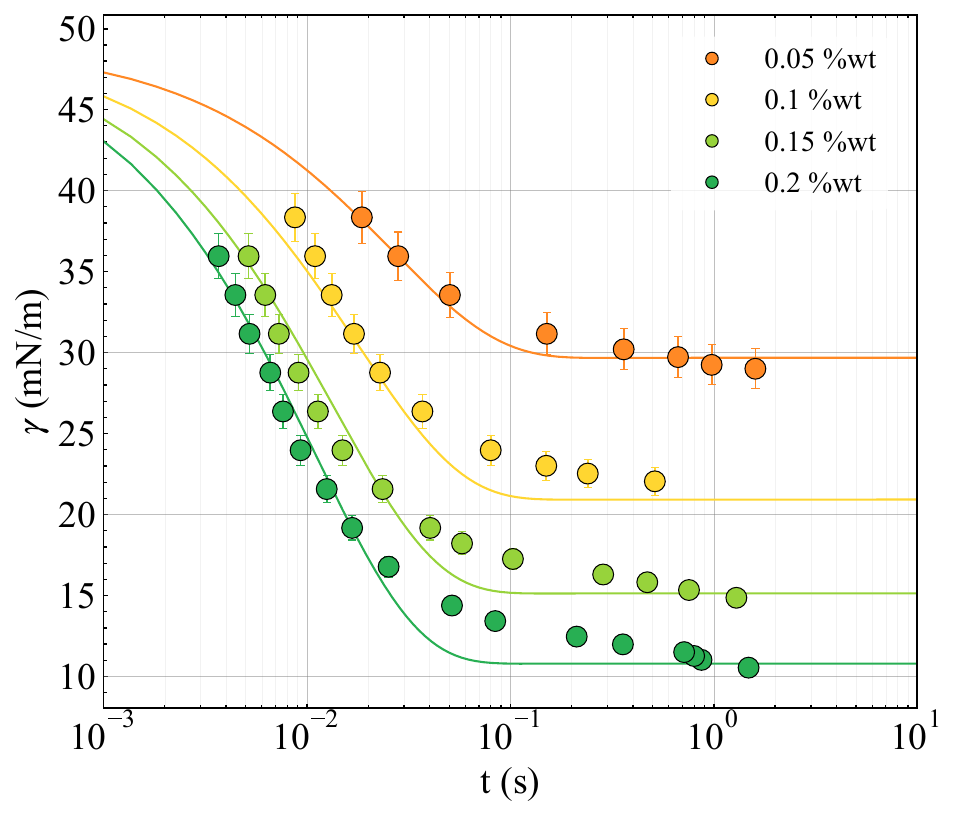}
    \caption{\textbf{Adsorption-controlled dynamics of SDS in the EDGE geometry.} Dynamic interfacial tension measured for SDS concentrations ranging from 0.05\% wt (0.2 CMC) to 0.2\% wt (0.8 CMC) (symbols). Solid lines are simultaneous fits of the adsorption-limited model, using the values from the equilibrium fit (see Table 2 from main text), yielding $k_a$= \num{9.63} \si{\cubic \meter \per \mol \per \second} and $k_d$ = \num{1.86} \si{\per \second}, with an overall RMS error of \num{1.32} \si{\mN \per \meter}.}
    \label{fig:SDS_3}
\end{figure}

In the adsorption regime, i.e. when diffusion can be neglected, Equations 8,9 and 12 from the main text are used to calculate the dynamic interfacial tension. The same fitting process is used for SDS. \autoref{fig:SDS_3} shows the best-fit for an adsorption-controlled dynamics for SDS in the EDGE geometry. A systematic deviation from the experimental data, especially at higher concentration is observed. Together with the criterion given in the main text $\Pi$, a mixed model would describe the SDS dynamics better.  

\subsection{Mixed adsorption-diffusion regime}
If the adsorption-controlled condition is not satisfied, adsorption
depletes the liquid adjacent to the interface and the subsurface
concentration becomes lower than the bulk concentration,

\begin{equation}
c_s < c_b.
\end{equation}

The dynamics are then controlled jointly by molecular adsorption and
diffusive replenishment. We do not introduce an adjustable accumulation
volume or an additional storage length in the subsurface region. Instead,
$c_s$ is assumed to adjust quasi-steadily so that the surfactant flux
supplied from the bulk is balanced at each time by the adsorption flux:

\begin{equation}
k_m(c_b-c_s)
=
\Gamma_\infty
R_{\mathrm{ads}}^{\mathrm{N}}(c_s,\phi),
\label{eq:nonionic_mixed_balance}
\end{equation}

with

\begin{equation}
R_{\mathrm{ads}}^{\mathrm{N}}(c_s,\phi)
=
k_a c_s(1-\phi)-k_d\phi,
\qquad
\frac{\mathrm{d}\phi}{\mathrm{d}t}
=
R_{\mathrm{ads}}^{\mathrm{N}}(c_s,\phi).
\label{eq:nonionic_mixed_rate}
\end{equation}

For EDGE, the mass-transfer coefficient is fixed by the independently
determined diffusion coefficient and by the device geometry,

\begin{equation}
k_m
=
\frac{A_f}{A_i}\frac{D}{L_D}.
\label{eq:EDGE_km_kinetics}
\end{equation}

Combining the adsorption law with the interfacial flux balance gives

\begin{equation}
c_s(\phi)
=
\frac{
k_m c_b+\Gamma_\infty k_d\phi
}{
k_m+\Gamma_\infty k_a(1-\phi)
}.
\label{eq:nonionic_mixed_cs}
\end{equation}

Thus, $c_s$ is not an additional dynamic variable and no accumulation
equation is required. At each time, $c_s$ is calculated from the current
coverage using Eq.~\eqref{eq:nonionic_mixed_cs}, and the resulting value is
inserted into Eq.~\eqref{eq:nonionic_mixed_rate}. The mixed problem
therefore reduces to a single differential equation for $\phi(t)$.

When $k_m$ is large compared with the interfacial adsorption capacity,
Eq.~\eqref{eq:nonionic_mixed_cs} gives $c_s\simeq c_b$, and the
adsorption-controlled model is recovered. When adsorption consumes
surfactant faster than it is replenished, $c_s$ decreases below $c_b$ and
the measured relaxation becomes sensitive to both adsorption and
diffusion.

The quantities $D$, $L_D$, $A_f/A_i$, $K$, $\Gamma_\infty$, and all
interfacial thermodynamic parameters remain fixed in the mixed analysis.
The model therefore prevents an unknown transport resistance from being
incorporated into an apparent value of $k_a$.

\clearpage
\addcontentsline{toc}{section}{References}
\bibliographystyle{ieeetr}
\bibliography{Ref_SI.bib}